\documentclass[utf8]{FrontiersinHarvard} 

\usepackage{url,hyperref,lineno,microtype,subcaption,verbatim,amsmath,amsfonts,amssymb,graphicx}
\usepackage[onehalfspacing]{setspace}

\def\keyFont{\fontsize{8}{11}\helveticabold }
\def\firstAuthorLast{Y. S. Dimant} 
\def\Authors{Y. S. Dimant\,$^{*}$}
\def\Address{Boston University, Center for Space Physics, Boston, MA, United States}
\def\corrAuthor{Corresponding Author}

\def\corrEmail{dimant@bu.edu}

\begin{document}
\onecolumn
\firstpage{1}

\title[Deriving Improved Plasma Fluid Equations from Collisional Kinetic Theory]{Deriving Improved Plasma Fluid Equations from Collisional Kinetic Theory}

\author[\firstAuthorLast ]{\Authors} 
\address{} 
\correspondance{} 

\extraAuth{}

\maketitle

\begin{abstract}

\section{}

Developing a quantitative understanding of wave plasma processes in the lower
ionosphere requires a reasonably accurate theoretical description of the
underlying physical processes. For such highly collisional plasma environment
as the E-region ionosphere, kinetic theory represents the most accurate
theoretical description of wave processes. For the analytical treatment,
however, the collisional kinetic theory is extremely complicated and succeeds
only in a limited number of physical problems. To date, most research applied
oversimplified fluid models that lack a number of critical kinetic aspects, so
that the coefficients in the corresponding fluid equations are often accurate
only to an order of magnitude. This paper presents the derivation for the
highly collisional, partially magnetized case relevant to E-region conditions.
It provides a more accurate reduction of the ion and, especially, electron
kinetic equations to the corresponding 5-moment fluid equations by using a new
set of analytic approximations. This derivation results in more accurate
fluid-model set of equations appropriate for most E-region problems. The
results of this paper could be used for a routine practical analysis when
working with actual data. The improved equations can also serve as a basis for
more accurate plasma fluid computer simulations.

\tiny
\keyFont{ \section{Keywords:} E-region ionosphere, magnetized plasma, plasma-neutral collisions, kinetic theory, fluid equations, 5-moment description}
\end{abstract}

\section{Introduction}

At altitudes of the equatorial and high-latitude E-region ionospheres, the
ionosphere is highly collisional in such a way that ions are almost
demagnetized by their frequent collisions with the surrounding neutral
molecules while electrons still remain strongly magnetized. Strong DC electric
fields perpendicular to the geomagnetic field cause there electrojets and give
rise to plasma instabilities whose nonlinear development produces plasma
density irregularities observed by radars and rockets.

Developing a quantitative understanding of wave plasma processes in the lower
ionosphere requires an accurate theoretical description of the underlying
physical processes. For such dissipative environment, collisional plasma
kinetic theory represents the most accurate theoretical description of wave
processes. Particle-in-cell (PIC) simulations present the most advanced way to
apply the kinetic approach, but such massive computer simulations
\citep{Oppenheim:Ion04,Oppenheim:Large-scale08,
Oppenheim:Kinetic13,Oppenheim:Newly20} are usually quite costly. In many
cases, simple estimates and parameter dependencies provided by an analytic
approach will suffice. For the analytical treatment, however, the collisional
kinetic theory is extremely complicated and succeeds only in a limited number
of physical problems. To date, most research applied an oversimplified fluid
model that lacks many critical kinetic aspects. These models mostly apply to
weakly collisional conditions. The coefficients in the simple fluid equations
are often accurate only to an order of magnitude because they were not
obtained using the full kinetic theory. This paper presents the derivation of
improved fluid equations for the highly collisional, partially magnetized case
relevant to E-region conditions, starting from a more consistent kinetic
approach. It provides more accurate values for the fluid-model coefficients.

There are different approaches to analytical description of low-frequency
plasma processes in the E-region ionosphere, including both the kinetic theory
and fluid models. Traditionally, the kinetic theory of the FB instability
applied an oversimplified BGK collision operator \citep{Bhatnagar:Model54}.
This operator does not follow from an accurate Boltzmann collision operator,
but represents an artificial construct. It simplifies dramatically the
analytical treatment, satisfying the particle number conservation, as well as
the momentum and energy balances (albeit under certain conditions, see below).
This simplified approach is reasonably applicable to the description of the
heavy ions, but it is totally unacceptable to the description of the light
electrons \citep{Dimant:Kinetic95a}.

More accurate approaches to the kinetic description of electrons under
conditions of the E-region wave processes, such as the Farley-Buneman (FB)
instability, have been developed by a few research groups.
\citet{Stubbe:Concept90} modified the BGK terms to allow for the different
rates of the electron energy and momentum losses. This simple modification,
however, does not follow from the Boltzmann operator and cannot be trusted.
Later, two independent research groups developed more sophisticated and
accurate approaches. Kissack and collaborators
\citep{Kissack:Electron95,Kissack:Effect97,
Kissack:Thermal_I_08,Kissack:Thermal_II_08} applied Grad's method
\citep{Grad:Kinetic49,Rodbard:Combined95}, while
\citet{Dimant:Kinetic95a} used an expansion in Legendre polynomials with
respect to the angles in the velocity space
\citep{Gurevich:Nonlinear78,Allis:Semidivergence82}. The latter kinetic
approach has allowed the authors to predict a new electron thermal-driven
instability in the lower E/upper D regions
\citep{Dimant:Kinetic95b,Dimant:Kinetic95c}, which has been later explained in
terms of a much simpler fluid model \citep{Dimant:Physical97}. This effect has
been verified by others \citep{Robinson:Effects98,St-Maurice:Role00}. Later, a
similar thermal-instability process has been also suggested for ions
\citep{Kagan:Thermal00,Dimant:Ion04,Dimant:Unified23}.

This paper presents a consistent reduction of the ion and electron kinetic
equations to the 5-moment fluid equations by using a new set of analytic
approximations. This derivation results in a more accurate fluid model
appropriate for most E-region plasma problems.

The paper is organized as follows. Section~\ref{General kinetic treatment}
introduces the collisional kinetic equation and reviews the generic procedure
for obtaining the moment equations. The collisional parts there are not
specified and remain in the general integral form.
Section~\ref{Section: for ions} describes ion momentum equation~obtained using
the BGK collision model. The most important part is
section~\ref{Section: for electrons}. It derives low-frequency electron-fluid
equations using a kinetic theory based on the efficient isotropization of the
electron distribution function in the velocity space
\citep{Gurevich:Nonlinear78,Dimant:Kinetic95a}. This requires a more detailed
and sophisticated treatment.
Section~\ref{General kinetic approach for electrons} derives the moment
equations where the heat conductivity and frictional heating are given in
terms of a still unspecified small directional part of the velocity
distribution function. To illustrate major ideas of closing the derivation,
section~\ref{Constant collisional parameters} describes the simplest case of
the constant (i.e., velocity-independent) kinetic collision parameters.
Section~\ref{General case of velocity} presents the general results obtained
in detail in the Appendix. Compared to the simplest electron-fluid equations
from section~\ref{Constant collisional parameters}, the general momentum and
thermal-balance equations include a larger number of coefficients, as well as
additional heat-conductivity terms. By their look, the latter may appear
collisionless, but they have arisen exclusively due to the velocity dependence
of the kinetic electron-neutral collision frequency. All these results could
be used for a routine practical analysis when working with actual data. The
improved equations can serve as a basis for more accurate plasma fluid
computer simulations.

\section{General kinetic treatment}
\label{General kinetic treatment}

This section discusses a general approach to deriving the fluid model from the
kinetic theory for any plasma particles. To avoid confusion, throughout this
paper we will use the following nomenclature. We denote various kinds of
particles (charged or neutral) by Latin subscripts: $p$, $q$, etc., that stand
for electrons ($e$), ions of various kinds ($i$) and neutrals ($n$), while
denoting vector components by Greek subscripts: $\alpha$, $\beta$, etc.

Non-relativistic kinetics of charged particles of the kind $p$ with the
velocity $\vec{v}_{p}$ at a given location $\vec{r}$ and time $t$ is described
by the Boltzmann kinetic equation,%
\begin{equation}
\partial_{t}f_{p}+\nabla\cdot\left(  \vec{v}_{p}f_{p}\right)  +\partial
_{\vec{v}_{p}}\cdot\left[  \frac{q_{p}}{m_{p}}\left(  \vec{E}+\vec{v}%
_{p}\times\vec{B}\right)  f_{p}\right]  =\left(  \frac{df_{p}}{dt}\right)
_{\operatorname{col}}, \label{conservative}%
\end{equation}
where $f_{p}(\vec{v}_{p},\vec{r},t)$ is the single-particle velocity
distribution function. The left-hand side (LHS) of Eq.~(\ref{conservative})
describes the collisionless (Vlasov) dynamics of the $p$-species charged
particles in smoothed over many particles electric ($\vec{E}$) and magnetic
($\vec{B}$) fields (for simplicity, we ignore here a gravity force); $q_{p}$
and $m_{p}$ are the $p$-particle charge and mass, respectively. The LHS of
Eq.~(\ref{conservative}) is intentionally written in a conservative
(divergence) form which is more convenient for deriving the moment equations.

The right-hand side (RHS) of Eq.~(\ref{conservative}), term $(df_{p}%
/dt)_{\operatorname{col}}=\sum_{q}S_{pq}$, is the collisional operator
describing binary collisions of the $p$-particles with all available kinds of
charged and neutral particles denoted by $q$ (including the $p$-particles
themselves). In the general case, the partial components $S_{pq}$ represent
integral operators that involve products of $f_{p}(\vec{v}_{p},\vec{r},t)$ by
$f_{q}(\vec{v}_{q},\vec{r},t)$. The partial operator $S_{qq}$ is quadratically
nonlinear, while $S_{pq}$ with $p\neq q$ are linear with respect to $f_{p}$.

The binary collisions can be either elastic or inelastic. Elastic collisions
conserve the total kinetic energy, momentum and angular momentum of the
colliding pair. The corresponding partial collisional operator, $S_{pq}$, can
be described by the well-known Boltzmann collision integral
\citep{Shkarofsky:Particle66,Gurevich:Nonlinear78,Lifshitz:Physical81,
Schunk:Ionospheres09,Khazanov:Kinetic11}. During an inelastic collision of a
charged particle with a neutral particle, a fraction of the total kinetic
energy goes to excitation (de-excitation) of the neutral particle (or ion) or
to release of electrons via ionization. Inelastic processes in the lower
ionosphere often involve molecular dissociation, recombination with ions and
electron attachment, accompanied by photon radiation or absorption. The
complete kinetic description of all these processes is complicated. In many
cases, however, inelastic collisions are close to elastic and one can continue
using Boltzmann's integral with minor modifications
\citep{Gurevich:Nonlinear78,Shkarofsky:Particle66}. Kinetic
Eq.~(\ref{conservative}) with Boltzmann's collision integral \emph{per ce}
represents a significant simplification over the full multi-particle kinetics,
but it still remains quite difficult for a mathematical treatment and requires
further simplifications.

Being interested in fluid-model equations that follow from kinetic
Eq.~(\ref{conservative}), right below we review the conventional approach to
deriving equations for the lowest-order moments of the distribution function.
The material of this section will serve as a guide for more specific
derivations of the following sections.

The three lowest-order velocity moments include the $p$-species particle
density,
\begin{equation}
n_{p}(\vec{r},t)\equiv\int f_{p}d^{3}v_{p}, \label{n_alpha}%
\end{equation}
mean fluid velocity,
\begin{equation}
\vec{V}_{p}(\vec{r},t)\equiv\left\langle \vec{v}_{p}\right\rangle =\frac
{1}{n_{p}}\int\vec{v}_{p}f_{p}d^{3}v_{p}, \label{V_alpha}%
\end{equation}
and effective temperature,
\begin{equation}
T_{p}(\vec{r},t)=\frac{m_{p}}{3}\left\langle \left(  \vec{v}_{p}-\vec{V}%
_{p}\right)  ^{2}\right\rangle =\frac{m_{p}}{3n_{p}}\int\left(  \vec{v}%
_{p}-\vec{V}_{p}\right)  ^{2}f_{p}d^{3}v_{p}. \label{Temperatura}%
\end{equation}
The derivations below will also involve other velocity-averaged quantities
defined by%
\begin{equation}
\left\langle \cdots\right\rangle \equiv\frac{1}{n_{p}}\int\left(
\cdots\right)  f_{p}d^{3}v_{p}. \label{average}%
\end{equation}
Integrations in Eqs.~(\ref{n_alpha})--(\ref{average}) are performed over the
entire 3-D velocity space.

First, we consider the particle-number balance. Integrating
Eq.~(\ref{conservative}) over the particle velocities with $f_{p}\rightarrow0$
as $v_{p}\equiv|\vec{v}_{p}|\rightarrow\infty$, we easily obtain the
continuity equation for the $p$-particle fluid,%
\begin{equation}
\partial_{t}n_{p}+\nabla\cdot\left(  n_{p}\vec{V}_{p}\right)  =\int\left(
\frac{df_{p}}{dt}\right)  _{\operatorname{col}}d^{3}v_{p}.
\label{conti_general}%
\end{equation}
The RHS of Eq.~(\ref{conti_general}) includes various particle sources and
losses, like ionization, recombination and electron attachment. The collisions
between the charged particles of the same species usually conserve the average
particle number and hence do not contribute to the RHS of
Eq.~(\ref{conti_general}).

Second, we obtain the momentum-balance equation that involves the mean fluid
drift velocity, $\vec{V}_{p}$. Integrating Eq.~(\ref{conservative}) with the
weighting function $m_{p}\vec{v}_{p}$, for a given vector-component $\alpha$
of the momentum density, we obtain
\begin{align}
&  m_{p}\partial_{t}\left(  n_{p}V_{p_{\alpha}}\right)  +\sum_{\beta=1}%
^{3}\partial_{x_{\beta}}\mathbf{P}_{p_{\alpha\beta}}+m_{p}\sum_{\beta=1}%
^{3}\partial_{x_{\beta}}\left(  n_{p}V_{p_{\alpha}}V_{p_{\beta}}\right)
\nonumber\\
&  =q_{p}\left[  E_{\alpha}+\frac{1}{c}\left(  \vec{V}_{p}\times\vec
{B}\right)  _{\alpha}\right]  n_{p}+m_{p}\int v_{p_{\alpha}}\left(
\frac{df_{p}}{dt}\right)  _{\operatorname{col}}d^{3}v_{p},
\label{momen_promezh}%
\end{align}
where $\mathbf{P}_{p}$ is the total pressure tensor with vector components
defined as
\begin{equation}
\mathbf{P}_{p_{\alpha\beta}}\equiv m_{p}\int\left(  v_{p_{\alpha}%
}-V_{p_{\alpha}}\right)  \left(  v_{p_{\beta}}-V_{p_{\beta}}\right)
f_{p}d^{3}v_{p}. \label{pressure_tensor}%
\end{equation}
It combines the isotropic pressure, $P_{p}\delta_{mn}$ ($\delta_{mn}=1$ if
$m=n$ and $\delta_{mn}=0$ otherwise), $P_{p}=n_{p}T_{p}$, with the viscosity
tensor, $\mathbf{\Pi}_{p_{\alpha\beta}}\equiv\mathbf{P}_{p_{\alpha\beta}%
}-P_{p}\delta_{\alpha\beta}$. Equation~(\ref{momen_promezh}) includes momentum
changes due to various average forces and those caused by particle density
variations. To exclude the latter and separate the net effect of the total
force, we multiply Eq.~(\ref{conti_general}) by $m_{p}\vec{V}_{p}$ and
subtract the resultant equation from Eq.~(\ref{momen_promezh}). This yields
the conventional momentum-balance equation,
\begin{equation}
m_{p}n_{p}\ \frac{D_{p}\vec{V}_{p}}{Dt}=q_{p}n_{p}\left[  \vec{E}+\frac{1}%
{c}\left(  \vec{V}_{p}\times\vec{B}\right)  \right]  -\nabla\cdot
\mathbf{P}_{p}+\vec{R}_{p}, \label{momentum-general}%
\end{equation}
where $D_{p}/Dt\equiv\partial_{t}+\vec{V}_{p}\cdot\nabla$ is the convective
(also called substantial or material) derivative for the average $p$-particle
flow and
\begin{equation}
\vec{R}_{p}\equiv m_{p}\int\left(  \vec{v}_{p}-\vec{V}_{p}\right)  \left(
\frac{df_{p}}{dt}\right)  _{\operatorname{col}}d^{3}v_{p}. \label{R_alpha}%
\end{equation}
Here and below, the \textquotedblleft dot\textquotedblright-products of a
vector, $\vec{a}$, with a two-component tensor, $\mathbf{A}$, depending on the
multiplier order, denote vectors with the components $\left(  \vec{a}%
\cdot\mathbf{A}\right)  _{\alpha}\equiv\sum_{\beta=1}^{3}a_{\beta}%
\mathbf{A}_{\beta\alpha}$ or $\left(  \mathbf{A}\cdot\vec{a}\right)  _{\alpha
}\equiv\sum_{\beta=1}^{3}\mathbf{A}_{\alpha\beta}a_{\beta}$. The tensor
divergence, $\nabla\cdot\mathbf{P}_{p}=\nabla P_{p}+\nabla\cdot\mathbf{\Pi
}_{p}$, represents a vector which uses the obvious symmetry $\mathbf{P}%
_{p_{\alpha\beta}}=\mathbf{P}_{p_{\beta\alpha}}$ following from
Eq.~(\ref{pressure_tensor}). The RHS of Eq.~(\ref{momentum-general}) includes
all smooth forces acting on the average particle flow of the charged
particles, such as the total Lorentz force, pressure gradient, and total
friction, $\vec{R}_{p}$. The latter is associated with collisions of the given
$p$-particles with all other charged or neutral particles. It includes no
momentum exchange between the same-species particles because their mutual
collisions automatically conserve the total momentum, $\int\vec{v}_{p}%
S_{pp}d^{3}v_{p}=0$.

Third, to obtain the total energy-balance equation, we integrate
Eq.~(\ref{conservative}) with the weighting function $m_{p}v_{p}^{2}/2$ and
obtain
\begin{equation}
\partial_{t}\mathcal{E}_{p}+\nabla\cdot\int\frac{m_{p}v_{p}^{2}}{2}\ \vec
{v}_{p}f_{p}d^{3}v_{p}=\vec{j}_{p}\cdot\vec{E}+\frac{m_{p}}{2}\int v_{p}%
^{2}\left(  \frac{df_{p}}{dt}\right)  _{\operatorname{col}}d^{3}%
v_{p},\label{en_balance_prom}%
\end{equation}
where $\mathcal{E}_{p}$ is the $p$-species average kinetic-energy density and
$\vec{j}_{p}$ is their electric-current density,
\begin{equation}
\mathcal{E}_{p}\equiv\int\frac{m_{p}v_{p}^{2}}{2}\ \ f_{p}d^{3}v_{p}%
,\qquad\vec{j}_{p}\equiv q_{p}n_{p}\vec{V}_{p}.\label{Eji}%
\end{equation}
Before proceeding, we separate from the kinetic particle velocity $\vec{v}%
_{p}$ the mean drift velocity $\vec{V}_{p}$, so that (\ref{en_balance_prom})
becomes
\begin{align}
&  \partial_{t}\left[  n_{p}\left(  \frac{m_{p}V_{p}^{2}}{2}+\frac{3T_{p}}%
{2}\right)  \right]  \nonumber\\
&  +\nabla\cdot\left[  n_{p}\left(  \frac{m_{p}V_{p}^{2}}{2}+\frac{5T_{p}}%
{2}\right)  \vec{V}_{p}+\mathbf{\Pi}_{p}\cdot\vec{V}_{p}+\frac{n_{p}m_{p}}%
{2}\left\langle \left(  \vec{v}_{p}-\vec{V}_{p}\right)  ^{3}\right\rangle
\right]  \nonumber\\
&  =\vec{j}_{p}\cdot\vec{E}+\vec{V}_{p}\cdot\vec{R}_{p}+\frac{m_{p}}{2}%
\int\left(  \vec{v}_{p}-\vec{V}_{p}\right)  ^{2}\left(  \frac{df_{p}}%
{dt}\right)  _{\operatorname{col}}d^{3}v_{p}\nonumber\\
&  +\frac{m_{p}V_{p}^{2}}{2}\int\left(  \frac{df_{p}}{dt}\right)
_{\operatorname{col}}d^{3}v_{p}.\label{Energy_prom}%
\end{align}
where $\left(  \vec{v}_{p}-\vec{V}_{p}\right)  ^{3}=\left\vert \vec{v}%
_{p}-\vec{V}_{p}\right\vert ^{2}\left(  \vec{v}_{p}-\vec{V}_{p}\right)  $.
Equation~(\ref{Energy_prom}) describes dynamic variations of the total energy
density. It includes a part associated with the average fluid motion,
$n_{p}m_{p}V_{p}^{2}/2$, and the internal thermal energy, $n_{p}T_{p}$. To
extract the equation exclusively for the particle temperature, $T_{p}$, we
multiply Eq.~(\ref{conti_general}) by $\left(  m_{p}V_{p}^{2}/2+3T_{p}%
/2\right)  $, take the scalar product of Eq.~(\ref{momentum-general}) with
$\vec{V}_{p}$ and subtract the resultant two equations from
Eq.~(\ref{Energy_prom}). This yields%
\begin{align}
&  \frac{3n_{p}}{2}\ \frac{D_{p}T_{p}}{Dt}+n_{p}T_{p}\nabla\cdot\vec{V}%
_{p}+\mathbf{\Pi}_{p}\cdot\nabla\cdot\vec{V}_{p}+\nabla\cdot\left[
\frac{n_{p}m_{p}}{2}\left\langle \left(  \vec{v}_{p}-\vec{V}_{p}\right)
^{3}\right\rangle \right]  \nonumber\\
&  =\frac{m_{p}}{2}\int\left(  \vec{v}_{p}-\vec{V}_{p}\right)  ^{2}\left(
\frac{df_{p}}{dt}\right)  _{\operatorname{col}}d^{3}v_{p}+\left(  \frac
{m_{p}V_{p}^{2}}{2}-\frac{3T_{p}}{2}\right)  \int\left(  \frac{df_{p}}%
{dt}\right)  _{\operatorname{col}}d^{3}v_{p},\label{dT_alpha_gen}%
\end{align}
where $\mathbf{\Pi}_{p}\cdot\nabla\cdot\vec{V}_{p}\equiv\sum_{\alpha,\beta
=1}^{3}\mathbf{\Pi}_{p_{\alpha\beta}}\nabla_{\alpha}\vec{V}_{p_{\beta}}$. Note
that after this step the electric field has been eliminated from the energy
balance equation. This is a crucial step in deriving the proper form of the
frictional heating, as described below.

Typically, equations like Eq.~(\ref{dT_alpha_gen}) represent the final form
of the thermal-balance equation. These equations are most convenient for
calculations. In order to clarify the physical meaning of some terms, however,
it is helpful to recast Eq.~(\ref{dT_alpha_gen}) in a slightly different form.
Rewriting the continuity Eq.~(\ref{conti_general}) as%
\[
\frac{D_{p}n_{p}}{Dt}+n_{p}\nabla\cdot\vec{V}_{p}=\int\left(  \frac{df_{p}%
}{dt}\right)  _{\operatorname{col}}d^{3}v_{p},
\]
we recast the two first terms in the LHS\ of Eq.~(\ref{dT_alpha_gen}) as%
\begin{align}
\frac{3n_{p}}{2}\ \frac{D_{p}T_{p}}{Dt}+n_{p}T_{p}\nabla\cdot\vec{V}_{p} &
=\frac{3n_{p}}{2}\ \frac{D_{p}T_{p}}{Dt}-T_{p}\ \frac{D_{p}n_{p}}{Dt}%
+T_{p}\int\left(  \frac{df_{p}}{dt}\right)  _{\operatorname{col}}d^{3}%
v_{p}\nonumber\\
&  =n_{p}T_{p}\ \frac{D_{p}s_{p}}{Dt}+T_{p}\int\left(  \frac{df_{p}}%
{dt}\right)  _{\operatorname{col}}d^{3}v_{p},\label{adiabatic}%
\end{align}
where $s_{p}\equiv\ln(T_{p}^{3/2}/n_{p})=\ln(P_{p}^{3/2}/n_{p}^{5/2})$
represents the specific entropy of the $p$-species fluid
\citep{Braginskii:Transport65} (for a single-atomic gas, the adiabatic
coefficient $\gamma=5/3$). This recast allows interpreting $n_{p}T_{p}%
\nabla\cdot\vec{V}_{p}$ as the adiabatic heating (cooling) term. The two
remaining terms in the LHS of Eq.~(\ref{dT_alpha_gen}) describe the work
performed by viscous forces and the fluid heat conductance. All these
processes are collisionless.

All collisional processes in the thermal balance Eq.~(\ref{dT_alpha_gen}) are
described by its RHS. After rearranging the last term in Eq.~(\ref{adiabatic})
to the RHS\ of Eq.~(\ref{dT_alpha_gen}), the last term there becomes $\left(
m_{p}V_{p}^{2}/2-5T_{p}/2\right)  \int\left(  df_{p}/dt\right)
_{\operatorname{col}}d^{3}v_{p}$. All integral terms involving $\left(
df_{p}/dt\right)  _{\operatorname{col}}$ describe the frictional heating and
thermal inflows (outflows) associated with possible emergence (disappearance)
of $p$-particles as a result of ionization, recombination, etc. For the
general form of $\left(  df_{p}/dt\right)  _{\operatorname{col}}$, calculating
the frictional heating is not an easy task. Below we employ two different
kinds of further approximation: one is more appropriate for heavy
single-charged ions (Section~\ref{Section: for ions}), while the other is
suitable for light electrons (Section~\ref{Section: for electrons}).

Before proceeding further, we emphasize that in general no truncated chain of
moment equations is closed because starting from the momentum equation every
further moment equation involves higher-order moments. To allow the moment
equation chain to be rigorously truncated, the most appropriate is the
near-equilibrium case when the particle distribution function, along with its
small perturbations, remains reasonably close to Maxwellian
\citep{Dimant:Kinetic95a,Kissack:Electron95}. This case allows describing the
particle kinetics using a restricted number of spatially and temporarily
varying parameters, such as the particle density, temperature, and average
drift velocity (5-moment equations). In real situations, however, this is not
always the case, that is why inconsistencies in the fluid description often
happen \citep[e.g.,][]{Garcia-Colin:Inconsistency04,Velasco:Inconsistencies02}.
Higher-order sets of equations allow more serious deviations from Maxwellian,
but still with a restricted number of additional fluid parameters. Fluid
models that include restricted numbers of equations using approximate
closures, such as the 5 or 13-moment models \citep{Schunk:Ionospheres09}, can
be successfully employed in situations when there are no sharp gradients,
extreme fields, abundant superthermal particles, or extremely large
temperature differences between different species of the colliding particles.
These conditions are usually met at the equatorial electrojet. If they cannot
be met, then the adequate description of plasma dynamics may require a direct
solution of the corresponding collisional kinetic equation.

\section{BGK collision kinetics and the fluid model for
ions}
\label{Section: for ions}

In this section, we consider the ion kinetics in low-ionized, highly
collisional plasma. Bearing in mind mainly the lower-E/upper-D regions of the
ionosphere or similar media, we can neglect coulomb collisions between the
charged particles compared to their much more frequent collisions with the
neutrals. For the ion-neutral collision integral, we can use the simple
Bhatnagar-Gross-Krook (BGK) model \citep{Bhatnagar:Model54}.
Disregarding ionization-recombination processes, we can write the simplest BGK
collision operator in the neutral frame of reference as%
\begin{equation}
\left(  \frac{df_{i}}{dt}\right)  _{\operatorname{col}}^{\mathrm{BGK}}=\nu
_{i}\left(  f_{\mathrm{eff}}-f_{i}\right)  , \label{S_in}%
\end{equation}
where $f_{\mathrm{eff}}$ is a fictitious Maxwellian function, normalized to
the locally varying ion density, $n_{i}(\vec{r},t)$, with the constant neutral temperature $T_{n}$:
\begin{equation}
f_{\mathrm{eff}}(\vec{v}_{i},\vec{r},t)\equiv n_{i}(\vec{r},t)\left(
\frac{m_{i}}{2\pi T_{n}}\right)  ^{3/2}\exp\left(  -\ \frac{m_{i}v_{i}^{2}%
}{2T_{n}}\right)  . \label{T_eff}%
\end{equation}
For the BGK model, it is essential that the ion-neutral collision frequency,
$\nu_{i}$, is assumed constant. The model collision term in the form of
Eq.~(\ref{S_in}) conserves the local number of particles. Applied to both ions
and neutrals, the BGK model also conserves the total momentum of the two
colliding particles.

The BGK model does not follow from Boltzmann's collision integral under any
rigorous approximations, but it is a reasonable fit for single-charged ions
that collide, predominantly elastically, with the surrounding neutrals of the same
(or close) mass. Recent 2-D hybrid computer simulations of the Farley-Buneman
instability that used for ions this kinetic equation \citep{Kovalev:Modeling08}
have demonstrated a good agreement with similar results of the more accurate
fully-kinetic PIC or hybrid simulations \citep{Janhunen:Recent95,
Oppenheim:Large-scale08,Oppenheim:Saturation96,Oppenheim:Hybrid95,Oppenheim:Ion04}.
There are two major reasons why this oversimplified model works well for the
predominantly ion-neutral collisions typical for the lower ionosphere. First,
within a 1000~K temperature range, the ion-neutral collision frequency is
nearly velocity-independent (Maxwell molecule collisions)
\citep{Schunk:Ionospheres09}. Second, collisions of ions with neutral particles
of the same or close mass have roughly equal rates of the average momentum and
energy transfer, described by the single parameter $\nu_{i}$. Both these
factors distinguish dramatically the ion-neutral collisions from the
electron-neutral ones, as we discuss in the following section.

For the distribution function of single-charged positive ions, $f_{i}(\vec
{v}_{i},\vec{r},t)$, the BGK kinetic equation in the conservative (divergence)
form is given by%
\begin{equation}
\partial_{t}f_{i}+\nabla\cdot\left(  \vec{v}_{i}f_{i}\right)  +\partial
_{\vec{v}_{i}}\cdot\left[  \left(  \frac{e\vec{E}}{m_{i}}+\Omega_{i}\vec
{v}_{i}\times\hat{b}\right)  f_{i}\right]  =\nu_{i}\left(  f_{\mathrm{eff}%
}-f_{i}\right)  . \label{conserva_ion}%
\end{equation}
In this section, we derive the 5-moment ion fluid-model equations for
$n_{i}=\int f_{i}d^{3}v_{i}$, $\vec{V}_{i}=\left\langle \vec{v}_{i}%
\right\rangle =\int\vec{v}_{i}f_{i}d^{3}v_{i}$, and $T_{i}=m_{i}\left\langle
\delta\vec{v}_{i}\right\rangle ^{2}/3=(m_{i}/3n_{i})\int\delta v_{i}^{2}%
f_{i}d^{3}v_{i}$, where $\delta\vec{v}_{i}\equiv\vec{v}_{i}-\vec{V}_{i}$. We
will largely follow the major steps of the previous section, but will do this
specifically for the ions in the BGK\ approximation. This will help us to
better understand the physical meaning of the algebraic manipulations and
intermediate results.

Integrating Eq.~(\ref{conserva_ion}) over velocities, we obtain the ion
continuity equation,%
\begin{equation}
\partial_{t}n_{i}+\nabla\cdot\left(  n_{i}\vec{V}_{i}\right)  =\frac
{D_{i}n_{i}}{Dt}+n_{i}\nabla\cdot\vec{V}_{i}=0. \label{ion_continuity}%
\end{equation}
\qquad

Further, integrating Eq.~(\ref{conserva_ion}) over ion velocities with the
weighting function $\vec{v}_{i}$, we obtain the total momentum balance
equation,%
\begin{equation}
\partial_{t}\left(  n_{i}\vec{V}_{i}\right)  +\partial_{x_{m}}\int v_{i_{k}%
}v_{i_{m}}fd^{3}v_{i}-n_{i}\left(  \frac{e\vec{E}}{m_{i}}+\Omega_{i}\vec
{V}_{i}\times\hat{b}\right)  =-\nu_{i}n_{i}\vec{V}_{i}.
\label{ion_momentum_total}%
\end{equation}
Using Eq.~(\ref{ion_continuity}) and recalling the definition of the pressure
in Eq.~(\ref{pressure_tensor}), we obtain the final momentum equation,%
\begin{equation}
m_{i}n_{i}\ \frac{D_{i}\vec{V}_{i}}{Dt}=m_{i}n_{i}\left(  \frac{e\vec{E}%
}{m_{i}}+\Omega_{i}\vec{V}_{i}\times\hat{b}\right)  -\nabla\cdot\mathbf{P}%
_{i}-\nu_{i}n_{i}m_{i}\vec{V}_{i}. \label{momentum_i}%
\end{equation}

Integrating Eq.~(\ref{conserva_ion}) with the weighting function $m_{i}%
v_{i}^{2}/2$, we obtain
\begin{align}
&  \partial_{t}\left(  \frac{m_{i}n_{i}V_{i}^{2}}{2}+\frac{3n_{i}T_{i}}%
{2}\right)  +\nabla\cdot\int\vec{v}_{i}f_{i}\ \frac{m_{i}v_{i}^{2}}{2}%
\ d^{3}v_{i}\nonumber\\
&  =n_{i}e\vec{E}\cdot\vec{V}_{i}+\nu_{i}\left[  \frac{3}{2}\ n_{i}\left(
T_{n}-T_{i}\right)  -\frac{m_{i}n_{i}V_{i}^{2}}{2}\right]  .
\label{ion_energy_total}%
\end{align}
Using the easily derivable relation%
\[
\int\vec{v}_{i}f_{i}\ \frac{m_{i}v_{i}^{2}}{2}\ d^{3}v_{i}=\vec{V}_{i}\left(
\frac{m_{i}n_{i}V_{i}^{2}}{2}+\frac{5n_{i}T_{i}}{2}\right)  +\vec{q}_{i},
\]
where the vector components of $\vec{q}_{i}$ are given by%
\begin{equation}
q_{i\alpha}\equiv m_{i}\int\delta v_{i_{\alpha}}\ \frac{\delta v_{i}^{2}}%
{2}\ f_{i}d^{3}v_{i}+\sum_{\beta=1}^{3}V_{i_{\beta}}\mathbf{\Pi}%
_{i_{\alpha\beta}}, \label{q_alpha}%
\end{equation}
we obtain%
\begin{align}
&  \frac{D_{i}}{Dt}\left(  \frac{m_{i}n_{i}V_{i}^{2}}{2}+\frac{3n_{i}T_{i}}%
{2}\right) \nonumber\\
&  +\left(  \frac{m_{i}n_{i}V_{i}^{2}}{2}+\frac{3n_{i}T_{i}}{2}\right)
\nabla\cdot\vec{V}_{i}+\nabla\cdot\left(  n_{i}T_{i}\vec{V}_{i}+\vec{q}%
_{i}\right) \nonumber\\
&  =n_{i}e\vec{E}\cdot\vec{V}_{i}+\nu_{i}\left[  \frac{3}{2}\ n_{i}\left(
T_{n}-T_{i}\right)  -\frac{m_{i}n_{i}V_{i}^{2}}{2}\right]  .
\label{ion_promka}%
\end{align}
Taking a scalar product of Eq.~(\ref{momentum_i}) with $\vec{V}_{i}$, we
obtain%
\begin{equation}
n_{i}\ \frac{D_{i}}{Dt}\left(  \frac{m_{i}V_{i}^{2}}{2}\right)  =n_{i}e\vec
{E}\cdot\vec{V}_{i}-\vec{V}_{i}\cdot\nabla\cdot\mathbf{P}_{i}-\nu_{i}%
m_{i}n_{i}V_{i}^{2}. \label{promezhutok_i}%
\end{equation}
Using Eq.~(\ref{ion_continuity}) and $\mathbf{P}_{i}=n_{i}T_{i}\mathbf{I}%
+\mathbf{\Pi}_{i}$, we rewrite Eq.~(\ref{promezhutok_i}) as%
\begin{align}
&  \frac{D_{i}}{Dt}\left(  \frac{n_{i}m_{i}V_{i}^{2}}{2}\right) \nonumber\\
&  =n_{i}e\vec{E}\cdot\vec{V}_{i}-\vec{V}_{i}\cdot\nabla\left(  n_{i}%
T_{i}\right)  -\vec{V}_{i_{\alpha}}\nabla_{\beta}\mathbf{\Pi}_{i_{\alpha\beta
}}-\nu_{i}m_{i}n_{i}V_{i}^{2}-\frac{n_{i}m_{i}V_{i}^{2}}{2}\ \nabla\cdot
\vec{V}_{i}. \label{average_dt}%
\end{align}
Rewriting Eq.~(\ref{ion_continuity}) as
\begin{equation}
T_{i}\ \frac{D_{i}}{Dt}\left(  \frac{3n_{i}}{2}\right)  +\frac{3n_{i}T_{i}}%
{2}\ \nabla\cdot\vec{V}_{i}=0 \label{tempuh}%
\end{equation}
we subtract both Eqs.~(\ref{average_dt}) and (\ref{tempuh}) from
Eq.~(\ref{ion_promka}) and obtain%
\begin{align*}
&  \frac{3n_{i}}{2}\ \frac{D_{i}T_{i}}{Dt}+n_{i}T_{i}\cdot\nabla\vec{V}%
_{i}+\nabla_{\alpha}q_{\alpha}-V_{i\beta}\nabla_{\alpha}\mathbf{\Pi
}_{i_{\alpha\beta}}\\
&  =\frac{\nu_{i}m_{i}n_{i}V_{i}^{2}}{2}+\frac{3}{2}\ \nu_{i}n_{i}\left(
T_{n}-T_{i}\right)  .
\end{align*}
Using%
\[
\nabla_{\alpha}q_{i\alpha}-V_{i\beta}\nabla_{\alpha}\mathbf{\Pi}%
_{i_{\alpha\beta}}=m_{i}\nabla_{\alpha}\int\delta v_{i_{\alpha}}\ \frac{\delta
v_{i}^{2}}{2}\ f_{i}d^{3}v_{i}+\mathbf{\Pi}_{i_{\alpha\beta}}\nabla_{\alpha
}V_{i\beta}%
\]
and $n_{i}\nabla\cdot\vec{V}_{i}=-D_{i}n_{i}/dt$, we obtain the final ion balance equation
\begin{align}
&  \frac{3n_{i}}{2}\ \frac{D_{i}T_{i}}{Dt}-T_{i}\ \frac{D_{i}n_{i}}{Dt}%
+\nabla\cdot\int m_{i}\delta\vec{v}_{i}\ \frac{\delta v_{i}^{2}}{2}%
\ f_{i}d^{3}v_{i}+\mathbf{\Pi}_{i_{\alpha\beta}}\nabla_{\alpha}V_{i\beta
}\nonumber\\
&  =\frac{\nu_{i}m_{i}n_{i}V_{i}^{2}}{2}+\frac{3}{2}\ \nu_{i}n_{i}\left(
T_{n}-T_{i}\right)  . \label{ion_temperature_equation}%
\end{align}

The two last terms in the LHS of the ion thermal balance
Eq.~(\ref{ion_temperature_equation}) describe the thermal conduction. The
thermal flux given by the integral term and $\mathbf{\Pi}_{i_{\alpha\beta}}$
should be determined from higher-order moment equations. Alternatively,
Eq.~(\ref{ion_temperature_equation}) could be closed using proper
approximations like Maxwellian $f_{i}$.

The first term in the RHS of Eq.~(\ref{ion_temperature_equation}) describes
the total ion frictional heating. This term has originated from a subtraction
of the frictional energy loss described by
\[
\nu_{i}\left(  \int\frac{m_{i}v^{2}}{2}\ f_{i}d^{3}v_{i}-\frac{3n_{i}T_{i}}%
{2}\right)  =\frac{\nu_{i}m_{i}n_{i}V_{i}^{2}}{2},
\]
from the total work performed during the collisional braking of the ion fluid
as a whole, $\nu_{i}m_{i}n_{i}V_{i}^{2}$, given by the last term in
Eq.~(\ref{promezhutok_i}). This means that when the ion fluid is frictionally
braked only the half of the released energy heats the ions, while the other
half goes directly to the colliding neutrals. Incidentally, this is precisely
the total amount of frictional heating for ions and neutrals of the equal
mass, $m_{n}=m_{i}$, following from a more detailed kinetic theory
\citep{Schunk:Ionospheres09,Dimant:Ion04},%
\[
\frac{\nu_{i}m_{i}m_{n}n_{i}V_{i}^{2}}{m_{i}+m_{n}}=\frac{\nu_{i}m_{i}%
n_{i}V_{i}^{2}}{2}.
\]
Thus, the BGK model of ion-neutral collisions describes correctly the ion
frictional heating if the colliding partners have the same or close masses.
This means that one can successfully employ for ions moment equations
(\ref{tempuh})--(\ref{ion_temperature_equation}) under moderate conditions
when the ion distribution function is reasonably close to Maxwellian. Such
conditions mostly apply to the equatorial E region, rather than to the
high-latitude ionosphere.

\section{Collisional kinetics and the fluid model for
electrons}\
\label{Section: for electrons}

This section is the central piece of this paper. It derives the electron fluid
equations from an approximate, but rigorous, kinetic theory based on
characteristics of the actual physical conditions and wave processes in the
E-region ionosphere. For electrons, the oversimplified BGK collision model
(employed above for ions) can apply only to plasma processes whose
characteristic wave frequencies exceed substantially the electron collision
frequencies. However, for low-frequency processes in the highly collisional
E/D-region ionosphere, where the opposite condition usually holds
\citep[see][and references therein]{Dimant:Kinetic95a}, the electron BGK collision
model is totally unsuitable. The main reason is that the rate of
electron-neutral collisional exchange of momentum, $\nu_{en}$, is a few orders
of magnitude larger than the corresponding rate of the energy exchange,
$\delta_{en}\nu_{en}$ \citep{Gurevich:Nonlinear78}. This means that during
collisions with heavy neutrals the light electrons scatter over angles in the
velocity space much more frequently than they change their kinetic energy. In
low-frequency processes of the lower ionosphere, this leads to an efficient
isotropization of the electron distribution function. The BGK model, however,
completely ignores this feature. In addition, the BGK\ model does not cover
the clearly pronounced velocity dependence of the kinetic electron-neutral
collision frequency $\nu_{en}(v_{e})$
\citep{Gurevich:Nonlinear78,Schunk:Ionospheres09}. This velocity dependence
plays an important role for some E-region instabilities,
\citep[see, e.g.,][and references therein]{Dimant:Physical97}, and it modifies the
instability and wave characteristics.

\subsection{General kinetic approach and momentum
equations}
\label{General kinetic approach for electrons}

In a weakly ionized plasma of the lower ionosphere, collisions of an electron
with other charged particles, including other electrons, $\nu_{ee},\nu_{ei}$,
are usually negligible compared to electron neutral collisions, $\nu_{e}\approx\nu_{en}$.
At altitudes above 75~km, strongly magnetized electrons, involved
in low-frequency processes with $\omega\ll\nu_{e}\ll\Omega_{e}$, have an
almost isotropic velocity distribution whose speed dependence can deviate
significantly from Maxwellian. For such processes, an adequate kinetic
description is by expanding the velocity distribution function $f_{e}(\vec
{r},t,\vec{v}_{e})$ in Legendre polynomials with respect to angles in the
velocity space
\citep{Shkarofsky:Particle66,Gurevich:Nonlinear78,Khazanov:Kinetic11}. To the
first-order accuracy with respect to a small anisotropy of $f_{e}(\vec
{r},t,\vec{v}_{e})$, one can represent the total electron distribution
function as a combination of the major isotropic part, $F_{0}(\vec{r}%
,t,v_{e})$, where $v_{e}\equiv|\vec{v}_{e}|$, and a relatively small
directional part determined by a single vector-function $\vec{f}_{1}(\vec
{r},t,v_{e})$ \citep{Gurevich:Nonlinear78,Dimant:Kinetic95a},
\begin{equation}
f_{e}(\vec{r},t,\vec{v}_{e})\approx F_{0}(\vec{r},t,v_{e})+\frac{\vec{f}%
_{1}(\vec{r},t,v_{e})\cdot\vec{v}_{e}}{v_{e}}=F_{0}+|\vec{f}_{1}|\cos\theta,
\label{f_approx}%
\end{equation}
where $\theta$ is the angle between $\vec{f}_{1}$ and $\vec{v}_{e}$. Here we
assume that $|\vec{f}_{1}|\ll F_{0}$, along with the similar inequalities for
the speed derivatives (see below). The major isotropic part, $F_{0}$,
determines scalar velocity-averaged characteristics of the electron fluid,
such as the electron density and temperature, while the small directional
part, $\vec{f}_{1}$, determines vector characteristics, such as the average
drift velocity and various fluxes. In this approximation, any higher-order
anisotropies are neglected. For electrons in the highly collisional E-region
ionosphere, the higher-order anisotropies usually play no role (see below).

The assumption of $|\vec{f}_{1}|\ll F_{0}$ is well justified for electrons
within the kinetic energy range $\mathcal{E}_{e}<2$\ eV ($v_{e}<1000$ km/s).
This range usually includes both the thermal bulk of electrons ($\mathcal{E}%
_{e}\lesssim0.03$~eV for the cold E-region ionosphere) and a significant
fraction of superthermal electrons. In this energy rate, the ratio of the
mean, mostly inelastic, collisional energy loss to that of the predominantly
elastic momentum loss, $\delta_{en}(v_{e})\nu_{en}(v_{e})/\nu_{en}%
(v_{e})=\delta_{en}(v_{e})$, is usually quite small: $\delta_{en}(v_{e}%
)\sim(2$--$4)\times10^{-3}$ \citep{Gurevich:Nonlinear78} (although it is two
orders of magnitude larger than the corresponding purely elastic rate,
$\delta_{en}^{\mathrm{elas}}\approx2m_{e}/m_{n}$). The ratio of $|\vec{f}%
_{1}|$ to $F_{0}$ is typically $\sim\sqrt{\delta_{en}}$, so that the
directional part of the electron distribution function in Eq.~(\ref{f_approx})
turns out to be automatically small compared to the major isotropic part,
$|\vec{f}_{1}|\ll F_{0}$. However, this brings up the following question. If
there were an imposed DC electric field, $\vec{E}\perp\vec{B}$, so strong that
the corresponding $\vec{E}\times\vec{B}$-drift velocity, $\vec{V}%
_{\mathrm{dr}}=\vec{E}\times\vec{B}/B^{2}$, would be comparable to the mean
electron thermal speed, $v_{Th}^{e}=(T_{e}/m_{e})^{1/2}$, then the condition
of $|\vec{f}_{1}|\ll F_{0}$ would become invalid. As a matter of fact,
however, such strong field would heat electrons so much that the heated
thermal velocity $v_{Th}^{e}$ would automatically exceed $\vec{V}_{\mathrm{dr}}$.
If the new electron temperature is $\lesssim23,000~$K (corresponding to 2~eV)
then the approximation (\ref{f_approx}) still holds. This is a significant difference
of electrons from heavy ions with $\delta_{in}\simeq 1$.

The kinetic equation with the general electron-neutral collision operator,
$(df_{e}/dt)_{\text{\textrm{coll}}}$, leads to the two coupled equations for
$F_{0}(\vec{r},t,v_{e})$ and $\vec{f}_{1}(\vec{r},t,v_{e})$
\citep{Gurevich:Nonlinear78,Dimant:Kinetic95a}:
\begin{subequations}
\label{dF_01}%
\begin{align}
\partial_{t}F_{0}+\frac{v_{e}}{3}\ \nabla\cdot\vec{f}_{1}-\frac{e}{3m_{e}%
v_{e}^{2}}\ \frac{\partial}{\partial v_{e}}\left(  v_{e}^{2}\vec{E}\cdot
\vec{f}_{1}\right)   &  =S_{0},\label{dF_0}\\
\partial_{t}\vec{f}_{1}-\Omega_{e}\hat{b}\times\vec{f}_{1}+v_{e}\nabla
F_{0}-\frac{e\vec{E}}{m_{e}}\ \frac{\partial F_{0}}{\partial v_{e}}  &
=\vec{S}_{1}, \label{dF_1}%
\end{align}
where%
\end{subequations}
\begin{equation}
S_{0}\equiv\frac{1}{2}\int_{-1}^{1}\left(  \frac{df_{e}}{dt}\right)
_{\operatorname{col}}d\left(  \cos\vartheta\right)  ,\qquad\vec{S}_{1}%
\equiv\frac{3}{2}\int_{-1}^{1}\left(  \frac{df_{e}}{dt}\right)
_{\operatorname{col}}\frac{\vec{f}_{1}}{\left\vert \vec{f}_{1}\right\vert
}\left(  \cos\vartheta\right)  d\left(  \cos\vartheta\right)  \label{S_0,S_1}%
\end{equation}
(note that the expressions for $S_{0,1}$ in \citet{Dimant:Kinetic95a} missed
the correct normalization factors). Bearing in mind moderately fast wave
processes, $\tau_{\mathrm{rec}}^{-1}\ll\omega\ll\nu_{e}$, where $\tau
_{\mathrm{rec}}$ is an effective recombination lifetime at a given altitude,
we will ignore ionization-recombination processes, as we did above for the
ions. The kinetic description of electrons based on Eq.~(\ref{dF_01}) differs
dramatically from any kinetic description based on the BGK collision model.

Fluid equations based on Eq.~(\ref{dF_01}), usually implying a
nearly-Maxwellian velocity distribution, have been successfully explored by a
number of researchers
\citep[see, e.g.,][and references therein]{Gurevich:Nonlinear78,Dimant:Kinetic95a}. However,
the form of major fluid equations presented in \citet{Gurevich:Nonlinear78},
Chapter 5, does not clearly show the basic structure of generic
Eqs.~(\ref{momentum-general}) and (\ref{dT_alpha_gen}) or similar ion
Eqs.~(\ref{momentum_i}) and (\ref{ion_temperature_equation}). By this, we mean
that Gurevich's equations show neither explicit adiabatic heating and cooling,
nor frictional heating $\propto V_{e,i}^{2}$. Adiabatic terms proportional to
$(\gamma_{e,i}-1)$ in \citet{Gurevich:Nonlinear78} Eqs.~(5.3) and (5.4)
and the corresponding terms in the following equations appear to have been
introduced \textquotedblleft by hand\textquotedblright\ and are actually
extraneous. One can verify that these adiabatic terms have already been
implicitly distributed among other terms of the temperature balance equations
within the corresponding fluxes given by \citet{Gurevich:Nonlinear78} equations
(5.8)--(5.11), so that they are accounted in Eqs.~(5.3)--(5.4) twice.

The explicit adiabatic terms show up naturally in the kinetic approaches based
on small perturbations of the distribution function shifted by the average
particle drift velocity. These approaches clearly differ from that based on
perturbations of the non-shifted velocity distribution, as in
Eq.~(\ref{f_approx}) resulting in Eq.~(\ref{dF_01}). For relatively small
drift velocities, however, the two different approaches should yield the same
results. Below we demonstrate that the kinetic approach based on
Eqs.~(\ref{f_approx})--(\ref{dF_01}) does reproduce in a rigorous and natural
way the electron fluid equations with the correct adiabatic heating and
cooling, frictional heating, etc. We will also calculate kinetic corrections
associated with the general velocity dependence of the electron-neutral
collision frequency and non-Maxwellian velocity distribution. The Appendix
contains details of these calculations.

In accord with the low-frequency condition of $\omega\ll\nu_{e}\ll\Omega_{e}$,
we neglect in Eq.~(\ref{dF_1}) the electron inertia term $\partial_{t}\vec
{f}_{1}$ and use a standard approximation $\vec{S}_{1}\approx-\nu_{e}%
(v_{e})\vec{f}_{1}$ \citep{Gurevich:Nonlinear78,Dimant:Kinetic95a}. This
allows us to close this set of equations in a simple way. As a result, we obtain
\begin{equation}
-\ \frac{e\vec{E}}{m_{e}}\ \frac{\partial F_{0}}{\partial v_{e}}-\Omega
_{e}\hat{b}\times\vec{f}_{1}+v_{e}\nabla F_{0}=-\nu_{e}(v_{e})\vec{f}_{1}.
\label{naprav}%
\end{equation}
Resolving this vector equation with respect to $\vec{f}_{1}$, we obtain%
\begin{equation}
\vec{f}_{1}(v_{e})=-\ \mathbf{N}(v_{e})\cdot\vec{K}F_{0}, \label{f_1_nach}%
\end{equation}
where the kinetic electron mobility tensor $\mathbf{N}(v)$ and the
differential vector operator $\vec{K}$ are given by%
\begin{equation}
\mathbf{N}(v_{e})\approx\left[
\begin{array}
[c]{ccc}%
\frac{\nu_{e}(v_{e})}{\Omega_{e}^{2}} & \frac{1}{\Omega_{e}} & 0\\
-\ \frac{1}{\Omega_{e}} & \frac{\nu_{e}(v_{e})}{\Omega_{e}^{2}} & 0\\
0 & 0 & \frac{1}{\nu_{e}(v_{e})}%
\end{array}
\right]  , \label{N}%
\end{equation}%
\begin{equation}
\vec{K}\equiv v_{e}\nabla-\frac{e\vec{E}}{m_{e}}\ \frac{\partial}{\partial
v_{e}}. \label{K}%
\end{equation}
Here and elsewhere, we neglect second-order small terms $\sim\nu_{e}^{2}$
compared to $\Omega_{e}^{2}$ and represent all tensors in the matrix form for
the Cartesian system $\hat{x},\hat{y},\hat{z}$ with the $\hat{z}$-axis along
$\vec{B}$. We can write Eqs.~(\ref{f_1_nach})--(\ref{N}) explicitly in terms
of the parallel ($\parallel$) and perpendicular ($\perp$) to $\vec{B}$
components as%
\begin{equation}
f_{1\parallel}=-\ \frac{1}{\nu_{e}(v_{e})}\ K_{\parallel}F_{0},\qquad\vec
{f}_{1\perp}=-\left(  \frac{\nu_{e}(v)\vec{K}_{\perp}}{\Omega_{e}^{2}}%
+\frac{\hat{b}}{\Omega_{e}}\times\vec{K}_{\perp}\right)  F_{0}, \label{f_1}%
\end{equation}
where $\hat{b}=\hat{z}$ is the unit vector along $\vec{B}$. The spatial
derivatives in Eq.~(\ref{f_1_nach}) or (\ref{f_1}) express the drift-diffusion
approximation in the collisional kinetic theory, while the velocity
derivatives describe electron energy variations caused by the electric field
$\vec{E}$.

Furthermore, expressing the RHS of Eq.~(\ref{dF_0}) as \citep{Gurevich:Nonlinear78}
\begin{equation}
S_{0}=\frac{1}{2v_{e}^{2}}\frac{\partial}{\partial v_{e}}\left[  v_{e}%
^{2}\delta_{en}\nu_{e}\left(  v_{e}F_{0}+\frac{T_{n}}{m_{e}}\frac{\partial
F_{0}}{\partial v_{e}}\right)  \right]  , \label{S_0_approx}%
\end{equation}
we obtain \citep{Dimant:Kinetic95a}
\begin{equation}
\partial_{t}F_{0}+\frac{1}{3v_{e}^{2}}\ \vec{K}\cdot\left(  v_{e}^{2}\vec
{f}_{1}\right)  =\frac{1}{2v_{e}^{2}}\frac{\partial}{\partial v_{e}}\left[
v_{e}^{2}\delta_{en}\nu_{e}\left(  v_{e}F_{0}+\frac{T_{n}}{m_{e}}%
\frac{\partial F_{0}}{\partial v_{e}}\right)  \right]  . \label{d_tF_0}%
\end{equation}
Expressing here $\vec{f}_{1}$ in terms of $F_{0}$ via Eq.~(\ref{f_1_nach}) or
(\ref{f_1}), we obtain a closed kinetic equation for the major isotropic
distribution function, $F_{0}(\vec{r},t,v)$. Its solution, with the use of
(\ref{f_1_nach}) or (\ref{f_1}), provides both parts of the distribution
function, so that its scalar and vector moments can be calculated by a
straightforward speed integration. According to Eq.~(\ref{f_approx}), the
lowest-order moments of the electron distribution function, such as the
electron density, mean drift velocity, and temperature, are given by
\begin{equation}
n_{e}\equiv4\pi\int_{0}^{\infty}\!\!F_{0}v_{e}^{2}dv_{e},\qquad\vec{V}%
_{e}\equiv\frac{4\pi}{3n_{e}}\int_{0}^{\infty}\!\!\vec{f}_{1}v_{e}^{3}%
dv_{e},\qquad T_{e}\equiv\frac{4\pi m_{e}}{3n_{e}}\int_{0}^{\infty}%
\!\!F_{0}v_{e}^{4}dv_{e}. \label{nTV}%
\end{equation}

A direct solution of the kinetic Eq.~(\ref{d_tF_0}) would be the most accurate
and general way to describing the electron behavior \citep{Dimant:Kinetic95a}. However, the goal of this
paper is to obtain a set of the lowest-order fluid equations in order to
properly describe E-region plasma processes, even if this set of equations is
not fully closed due to possible deviations of the electron velocity
distribution function from Maxwellian.

As above, we start from the particle conservation. Using the definitions of
Eq.~(\ref{nTV}) and integrating Eq.~(\ref{d_tF_0}) over $v_{e}$ with the
weighting function $4\pi v_{e}^{2}$, we obtain the standard electron
continuity equation,%
\begin{equation}
\partial_{t}n_{e}+\nabla\cdot(n_{e}\vec{V}_{e})=0. \label{e_conti}%
\end{equation}

Next, we integrate Eq.~(\ref{naprav}) with the weighting function $4\pi
v_{e}^{3}/(3n_{e})$. Applying the integration by parts, we obtain an equation%
\begin{equation}
\frac{e\vec{E}}{m_{e}}-\Omega_{e}\hat{b}\times\vec{V}_{e}+\frac{\nabla\left(
n_{e}T_{e}\right)  }{m_{e}n_{e}}+\frac{4\pi}{3n_{e}}\int_{0}^{\infty}\vec
{f}_{1}\nu_{e}v_{e}^{3}dv_{e}=0, \label{c_nachala}%
\end{equation}
that describes the momentum balance of the inertialess electron fluid.
Equation~(\ref{c_nachala}) includes the Lorentz force, pressure gradient, and
collisional friction. As we show in the Appendix, in the general case of a
velocity-dependent collision frequency, $\nu_{e}(v)$, the last term in the
LHS\ of Eq.~(\ref{c_nachala}), in addition to the collisional friction, may
also include an anisotropic addition to the total pressure gradient.

Taking a scalar product of Eq.~(\ref{c_nachala}) with $m_{e}n_{e}\vec{V}_{e}$,
we obtain the expression%
\begin{equation}
\vec{V}_{e}\cdot\left[  n_{e}e\vec{E}+\nabla\left(  n_{e}T_{e}\right)
\right]  +\frac{4\pi m_{e}}{3}\ \vec{V}_{e}\cdot\!\int_{0}^{\infty}\vec{f}%
_{1}\nu_{e}v_{e}^{3}dv_{e}=0 \label{direct_analog}%
\end{equation}
analogous to Eq.~(\ref{promezhutok_i}) for ions. This expression represents
the total work done by the electric field and other forces on the average
electron flow. We will employ this expression right below.

Now we derive an equation describing the total energy balance. Integrating
Eq.~(\ref{d_tF_0}) with the weighting function $2\pi m_{e}v_{e}^{4}$, we
obtain%
\begin{align}
&  \partial_{t}\left(  \frac{3n_{e}T_{e}}{2}\right)  +\frac{2\pi m_{e}}%
{3}\ \nabla\cdot\int_{0}^{\infty}\ \vec{f}_{1}v_{e}^{5}dv_{e}+n_{e}e\vec
{E}\cdot\vec{V}_{e}\nonumber\\
&  =-\ 2\pi m_{e}\int_{0}^{\infty}\left(  v_{e}F_{0}+\frac{T_{n}}{m_{e}}%
\frac{dF_{0}}{dv_{e}}\right)  \delta_{en}\nu_{e}v_{e}^{3}dv_{e}.
\label{electron_energy}%
\end{align}
Using Eq.~(\ref{direct_analog}), we eliminate from Eq.~(\ref{electron_energy})
the work done by the electric field on the average flow, $n_{e}e\vec{E}%
\cdot\vec{V}_{e}$, and obtain%
\begin{align}
&  \partial_{t}\left(  \frac{3n_{e}T_{e}}{2}\right)  -\vec{V}_{e}\cdot
\nabla\left(  n_{e}T_{e}\right)  +\frac{2\pi m_{e}}{3}\ \nabla\cdot\int%
_{0}^{\infty}\!\!\ \vec{f}_{1}v_{e}^{5}dv_{e}\nonumber\\
&  =\frac{4\pi m_{e}}{3}\ \vec{V}_{e}\cdot\int_{0}^{\infty}\!\!\vec{f}_{1}%
\nu_{e}v_{e}^{3}dv_{e}-2\pi m_{e}\int_{0}^{\infty}\!\!\left(  v_{e}F_{0}%
+\frac{T_{n}}{m_{e}}\frac{dF_{0}}{dv_{e}}\right)  \delta_{en}\nu_{e}v_{e}%
^{3}dv_{e}. \label{equivoka}%
\end{align}
Here we have rearranged the terms between the two sides of the equation in
such a way that all terms proportional to the collision frequency remain in
the RHS, while all other terms are put in the LHS. After so doing, it may be
tempting to interpret the first term in the RHS of Eq.~(\ref{equivoka})\ as
the electron frictional heating. In the general case of velocity-dependent
$\nu_{e}(v)$, however, this interpretation would not be perfectly accurate, as
we show in Appendix and section~\ref{General case of velocity} below.

Equation~(\ref{equivoka}) is not yet the final form of the thermal-balance
equation. It needs to be further transformed into a form similar to
Eq.~~(\ref{dT_alpha_gen}) or (\ref{ion_temperature_equation}). In Appendix, we
develop this recast for the general case of velocity-dependent $\nu_{e}(v)$.
However, right below we proceed with the simplest model of constant $\nu_{e}$
and $\delta_{en}$. This model is inaccurate for electron-neutral collisions of the lower ionosphere
\citep{Gurevich:Nonlinear78,Schunk:Ionospheres09}, but it will allow us to
clarify basic ideas of closing Eq.~(\ref{equivoka}).

\subsection{Constant collisional parameters}
\label{Constant collisional parameters}

For constant $\nu_{e}$ and $\delta_{en}$, using the definitions of
Eq.~(\ref{nTV}) and integrating the last term of Eq.~(\ref{equivoka}) by
parts, we obtain%

\begin{align}
&  \partial_{t}\left(  \frac{3n_{e}T_{e}}{2}\right)  -\vec{V}_{e}\cdot
\nabla\left(  n_{e}T_{e}\right)  +\frac{2\pi m_{e}}{3}\ \nabla\cdot\int%
_{0}^{\infty}\ \vec{f}_{1}v_{e}^{5}dv_{e}\nonumber\\
&  =m_{e}\nu_{e}n_{e}V_{e}^{2}+\frac{3}{2}\ \delta_{en}\nu_{e}n_{e}\left(
T_{n}-T_{e}\right)  , \label{promka}%
\end{align}
Using\ Eq.~(\ref{f_1_nach}), we rewrite the third term in the LHS as%
\begin{equation}
\frac{2\pi m_{e}}{3}\ \nabla\cdot\int_{0}^{\infty}\ \vec{f}_{1}v_{e}^{5}%
dv_{e}=-\ \frac{5}{2m_{e}}\ \nabla\cdot\mathbf{N\cdot}\left[  \nabla\left(
\lambda n_{e}T_{e}^{2}\right)  +n_{e}T_{e}e\vec{E}\right]  . \label{kuku}%
\end{equation}
Here the double-dot product involving a tensor means $\nabla\cdot
\mathbf{N}\cdot\nabla\ldots=\sum_{\alpha,\beta=1}^{3}\partial_{x_{\alpha}%
}(\mathbf{N}_{\alpha\beta}\partial_{x_{\beta}}\ldots)$ (and similar for
$\nabla\cdot\mathbf{N}\cdot\vec{E}$) and we have also introduced a
dimensionless parameter of order unity, $\lambda$,
\begin{equation}
\lambda\equiv\frac{4\pi m_{e}^{2}}{15n_{e}T_{e}^{2}}\int_{0}^{\infty}v_{e}%
^{6}F_{0}dv_{e}=\frac{m_{e}\int_{0}^{\infty}v_{e}^{6}F_{0}dv_{e}}{5T_{e}%
\int_{0}^{\infty}v_{e}^{4}F_{0}dv_{e}}=\frac{3\left(  \int_{0}^{\infty}%
F_{0}v_{e}^{2}dv_{e}\right)  \int_{0}^{\infty}F_{0}v_{e}^{6}dv_{e}}{5\left(
\int_{0}^{\infty}F_{0}v_{e}^{4}dv_{e}\right)  ^{2}}. \label{lambda}%
\end{equation}
Note that for the Maxwellian isotropic part of the electron distribution
function,%
\begin{equation}
F_{0}=n_{e}\left(  \frac{m_{e}}{2\pi T_{e}}\right)  ^{3/2}\exp\left(
-\ \frac{m_{e}v_{e}^{2}}{2T_{e}}\right)  , \label{Maxwellian}%
\end{equation}
we have $\lambda=1$.

Using Eq.~(\ref{nTV}) and (\ref{K}), we obtain%
\begin{equation}
\vec{V}_{e}=-\ \frac{4\pi}{3n_{e}}\ \mathbf{N}\cdot\int_{0}^{\infty}v_{e}%
^{3}\vec{K}F_{0}dv_{e}=-\mathbf{N}\cdot\left[  \frac{e\vec{E}}{m_{e}}%
+\frac{\nabla\left(  n_{e}T_{e}\right)  }{m_{e}n_{e}}\right]  .
\label{V_e_cherez}%
\end{equation}
Multiplying Eq.~(\ref{V_e_cherez}) by $m_{e}n_{e}T_{e}$, we can rewrite it as%
\[
-\mathbf{N}\cdot\left(  n_{e}T_{e}e\vec{E}\right)  =m_{e}n_{e}T_{e}\vec{V}%
_{e}+\mathbf{N}\cdot T_{e}\nabla\left(  n_{e}T_{e}\right)  .
\]
This relation allows us to eliminate the electric field from Eq.~(\ref{kuku}),
so that the latter becomes%
\begin{align}
&  \frac{2\pi m_{e}}{3}\ \nabla\cdot\int_{0}^{\infty}\ \vec{f}_{1}v_{e}%
^{5}dv_{e}\nonumber\\
&  =\frac{5}{2m_{e}}\ \nabla\cdot\left\{  \mathbf{N\cdot}\left[  \left(
1-\lambda\right)  T_{e}^{2}\nabla n_{e}-\left(  2\lambda-1\right)  n_{e}%
T_{e}\nabla T_{e}\right]  +m_{e}n_{e}T_{e}\vec{V}_{e}\right\}  .
\label{zakroma}%
\end{align}
Using Eqs.~(\ref{e_conti}) and (\ref{zakroma}), after a simple algebra,%
\begin{align}
&  \partial_{t}\left(  \frac{3n_{e}T_{e}}{2}\right)  -\vec{V}_{e}\cdot
\nabla\left(  n_{e}T_{e}\right)  +\frac{5}{2}\ \nabla\cdot\left(  n_{e}%
T_{e}\vec{V}_{e}\right) \nonumber\\
&  =\partial_{t}\left(  \frac{3n_{e}T_{e}}{2}\right)  -\vec{V}_{e}\cdot
\nabla\left(  n_{e}T_{e}\right)  +\frac{5}{2}\ \nabla\cdot\left(  n_{e}%
T_{e}\vec{V}_{e}\right) \nonumber\\
&  -\frac{5T_{e}}{2}\left[  \partial_{t}n_{e}+\nabla\cdot(n_{e}\vec{V}%
_{e})\right]  =\frac{3n_{e}}{2}\ \frac{D_{e}T_{e}}{Dt}-T_{e}\ \frac{D_{e}%
n_{e}}{Dt}, \label{simple_algebra}%
\end{align}
we obtain the sought-for temperature-balance equation in a more standard
form,
\begin{equation}
\frac{3n_{e}}{2}\ \frac{D_{e}T_{e}}{Dt}-T_{e}\ \frac{D_{e}n_{e}}{Dt}%
-\nabla\cdot\vec{q}_{e}=m_{e}\nu_{e}n_{e}V_{e}^{2}+\frac{3}{2}\ \delta_{en}%
\nu_{e}n_{e}\left(  T_{n}-T_{e}\right)  . \label{el_temper_constant_final}%
\end{equation}
Here the electron thermal flux density, $\vec{q}_{e}$, is given by%
\begin{equation}
\vec{q}_{e}=\frac{5T_{e}}{2m_{e}}\ \mathbf{N}\cdot\left[  \left(
2\lambda-1\right)  n_{e}\nabla T_{e}+\left(  \lambda-1\right)  T_{e}\nabla
n_{e}\right]  =\vec{q}_{e\parallel}+\vec{q}_{e\mathrm{P}}+\vec{q}%
_{e\mathrm{H}}, \label{q_e_stac}%
\end{equation}
where its explicit parallel, Pedersen, and Hall components are given by%
\begin{align}
\vec{q}_{e\parallel}  &  =\frac{5T_{e}\left[  \left(  2\lambda-1\right)
n_{e}\nabla_{\parallel}T_{e}+\left(  \lambda-1\right)  T_{e}\nabla_{\parallel
}n_{e}\right]  }{2m_{e}\nu_{e}},\nonumber\\
\vec{q}_{e\mathrm{P}}  &  =\frac{5T_{e}\nu_{e}\left[  \left(  2\lambda
-1\right)  n_{e}\nabla_{\perp}T_{e}+\left(  \lambda-1\right)  T_{e}%
\nabla_{\perp}n_{e}\right]  }{2m_{e}\Omega_{e}^{2}},\label{q_IIPH_stac}\\
\vec{q}_{e\mathrm{H}}  &  =\frac{\Omega_{e}}{\nu_{e}}\left(  \hat{b}\times
\vec{q}_{e\mathrm{P}}\right)  =\frac{5T_{e}\hat{b}\times\left[  \left(
2\lambda-1\right)  n_{e}\nabla_{\perp}T_{e}+\left(  \lambda-1\right)
T_{e}\nabla_{\perp}n_{e}\right]  }{2m_{e}\Omega_{e}}.\nonumber
\end{align}
The two first terms in the LHS of Eq.~(\ref{el_temper_constant_final}), as
well as the similar ones in Eq.~(\ref{dT_alpha_gen}) or
(\ref{ion_temperature_equation}), describe adiabatic heating or cooling of the
electron fluid, while $\nabla\cdot\vec{q}_{e}$ describes the heat
conductivity. Note that the Hall component of $\vec{q}_{e}$ can contribute
into electron heat conductance only as a quadratically-nonlinear effect and,
since $\nabla\cdot\vec{q}_{e\mathrm{H}}\propto(\nabla_{\perp}n_{e}\times
\nabla_{\perp}T_{e})$ only if the gradients of $n_{e}$ and $T_{e}$ are not parallel.

As mentioned above, for Maxwellian $F_{0}(v_{e})$ we have $\lambda=1$, so that
the term in $\vec{q}_{e}$ proportional to $\nabla n_{e}$ disappears. This fact
can be understood as follows. If the major part of the distribution function
remains Maxwellian then it is determined only by two space-dependent
parameters: the density, $n_{e}$, and the temperature, $T_{e}$. If there is a
density gradient but no temperature gradient, then electrons of all energies
will diffuse from denser regions to less dense ones with no redistribution of
the temperature and hence with no heat conductivity.

If the electron velocity distribution deviates from Maxwellian (this happens,
e.g., when a low-ionized plasma heated by strong electric fields is embedded
in an abundant cold neutral atmosphere with a significantly different
temperature \citep{Milikh:Model03}) then the situation is more complicated. The
effective electron temperature $T_{e}$, which is proportional to the mean
electron chaotic energy, can be uniformly distributed, but the details of the
electron energy distribution may differ significantly in different regions of
space. The energy transport is stronger for electrons with higher energies
than it is for lower-energy electrons. Hence, if there are spatial gradients
of high-energy distribution tails then more energetic particles provide
stronger energy redistribution. This may make, e.g., some less dense regions
to be on average more energetic than the denser regions, even if they had
initially equal effective temperatures. Moreover, it is even possible to
imagine a situation when electron heat is transferred from cooler regions to
hotter ones, leading to a further electron temperature elevation in the
latter. This counter-intuitive, but theoretically possible, effect should not
surprise because a strongly non-Maxwellian, i.e., strongly non-equilibrium
plasma cannot be adequately described by the conventional equilibrium thermodynamics.

\subsection{Velocity-dependent parameters}
\label{General case of velocity}

In the actual lower ionosphere, the electron-neutral kinetic collision
frequency, $\nu_{e}$, and the energy loss fraction, $\delta_{en}$, have
clearly pronounced velocity dependencies
\citep{Gurevich:Nonlinear78,Schunk:Ionospheres09}. This does not allow $\nu
_{e}(v_{e})$ and $\delta_{en}(v_{e})$ to be factored out from the integrals in
Eqs.~(\ref{c_nachala})--(\ref{equivoka}), making the derivation of the general
momentum and temperature-balance equations more complicated than that
described in section~\ref{Constant collisional parameters}. Such derivation is
developed in detail in the Appendix, while here we only present the results.

In the general case of velocity-dependent $\nu_{e}$ and $\delta_{en}$,
electron continuity Eq.~(\ref{e_conti}) stays the same. The other two moment
equations have the same basic structure as (\ref{V_e_cherez}) and
(\ref{el_temper_constant_final}), but they contain additional terms and
include a large number of dimensionless factors of order unity listed in
Eqs.~(\ref{alphy_again})--(\ref{xi}) below.

The general inertialess expression for the average electron drift velocity
$\vec{V}_{e}$ is given by%
\begin{equation}
\vec{V}_{e}=-\ \frac{1}{m_{e}}\ \mathbf{M}\cdot\left[  e\vec{E}+\frac
{\nabla_{\perp}\left(  n_{e}T_{e}\right)  }{n_{e}}+\frac{\beta_{\parallel}%
}{\alpha_{\parallel}}\ \frac{\nabla_{\parallel}\left(  n_{e}T_{e}\right)
}{n_{e}}\right]  , \label{final_momentum_V}%
\end{equation}
where%
\begin{equation}
\mathbf{M}\equiv\frac{4\pi}{3n_{e}}\int_{0}^{\infty}\frac{d\left(  v_{e}%
^{3}\mathbf{N}(v_{e})\right)  }{dv_{e}}\ F_{0}(v_{e})dv_{e}=\left[
\begin{array}
[c]{ccc}%
\frac{\alpha_{P}\left\langle \nu_{e}\right\rangle _{e}}{\Omega_{e}^{2}} &
\frac{1}{\Omega_{e}} & 0\\
-\ \frac{1}{\Omega_{e}} & \frac{\alpha_{P}\left\langle \nu_{e}\right\rangle
_{e}}{\Omega_{e}^{2}} & 0\\
0 & 0 & \alpha_{\parallel}\left\langle \frac{1}{\nu_{e}}\right\rangle _{e}%
\end{array}
\right]  , \label{M_e_again}%
\end{equation}
an the tensor $\mathbf{N}$ is given by Eq.~(\ref{N}) and $\left\langle
\cdots\right\rangle _{e}$ denotes the velocity average over the major
(isotropic) part of the electron distribution function,%
\begin{equation}
\left\langle \cdots\right\rangle _{e}=\frac{4\pi\int_{0}^{\infty}\left(
\cdots\right)  F_{0}(v_{e})v_{e}^{2}dv_{e}}{n_{e}}=\frac{\int_{0}^{\infty
}\left(  \cdots\right)  F_{0}(v_{e})v_{e}^{2}dv_{e}}{\int_{0}^{\infty}%
F_{0}(v_{e})v_{e}^{2}dv_{e}}. \label{<...>}%
\end{equation}
The electron current density is given by $\vec{j}_e=en_e\vec{V}_{e}$, so that the electric
conductivity tensor is given by $\boldsymbol{\sigma}_e = (n_e e^2/m_e)\mathbf{M}$. The corresponding diagonal terms
represent the Pedersen ($\propto \alpha_P$) and parallel ($\propto \alpha_{\parallel}$), while the antisymmetric
off-diagonal terms ($\propto 1/\Omega_e$) represent the Hall conductivity.

The general thermal-balance equation is given by
\begin{align}
&  \frac{3n_{e}}{2}\ \frac{D_{e}T_{e}}{Dt}-T_{e}\ \frac{D_{e}n_{e}}%
{Dt}\nonumber\\
&  +\frac{5}{2}\left(  \frac{\rho_{\parallel}-\beta_{\parallel}}%
{\alpha_{\parallel}}\right)  n_{e}T_{e}\nabla_{\parallel}\cdot\vec
{V}_{e\parallel}+\left(  \frac{5\rho_{\parallel}-3\alpha_{\parallel}%
-2\beta_{\parallel}}{2\alpha_{\parallel}}\right)  \vec{V}_{e\parallel}%
\cdot\nabla_{\parallel}\left(  n_{e}T_{e}\right)  -\nabla\cdot\vec{q}%
_{e}\nonumber\\
&  =\alpha_{P}\left\langle \nu_{e}\right\rangle _{e}m_{e}n_{e}\left(
V_{e\perp}^{2}+\frac{V_{\parallel}^{2}}{\alpha_{\parallel}\xi}\right)  -2\pi
m_{e}\int_{0}^{\infty}\!\!v_{e}^{3}\delta_{en}\nu_{e}\left(  v_{e}F_{0}%
+\frac{T_{n}}{m_{e}}\frac{dF_{0}}{dv_{e}}\right)  dv_{e},
\label{Thermal_balance_fin}%
\end{align}
where%
\begin{equation}
\vec{q}_{e}\equiv\vec{q}_{eP}+\vec{q}_{eH}+\vec{q}_{e\parallel}=\mathbf{X}%
\cdot\frac{\nabla T_{e}}{T_{e}}+\left(  \mathbf{X}-\mathbf{\Lambda}\right)
\cdot\frac{\nabla n_{e}}{n_{e}}, \label{qq_e}%
\end{equation}
is the thermal-flux density with
\begin{subequations}
\label{xxLL}%
\begin{align}
\mathbf{X}  &  =\frac{5n_{e}T_{e}}{2m_{e}}\left[
\begin{array}
[c]{ccc}%
\frac{\chi_{P}\left\langle \nu_{e}\right\rangle }{\Omega_{e}^{2}} & \frac
{\chi_{H}}{\Omega_{e}} & 0\\
-\ \frac{\chi_{H}}{\Omega_{e}} & \frac{\chi_{P}\left\langle \nu_{e}%
\right\rangle }{\Omega_{e}^{2}} & 0\\
0 & 0 & \frac{\chi_{\parallel}}{\left\langle \nu_{e}\right\rangle }%
\end{array}
\right]  ,\label{xx}\\
\mathbf{\Lambda}  &  =\frac{5n_{e}T_{e}}{2m_{e}}\left[
\begin{array}
[c]{ccc}%
\frac{\mu_{P}\left\langle \nu_{e}\right\rangle }{\Omega_{e}^{2}} &
\frac{\lambda}{\Omega_{e}} & 0\\
-\ \frac{\lambda}{\Omega_{e}} & \frac{\mu_{P}\left\langle \nu_{e}\right\rangle
}{\Omega_{e}^{2}} & 0\\
0 & 0 & \frac{\mu_{\parallel}}{\left\langle \nu_{e}\right\rangle }%
\end{array}
\right]  . \label{LL}%
\end{align}
The explicit Pedersen, Hall, and parallel components of $\vec{q}_{e}$ are
given by
\end{subequations}
\begin{align}
\vec{q}_{eP}  &  \equiv\frac{5T_{e}\left\langle \nu_{e}\right\rangle \left[
\chi_{P}n_{e}\nabla_{\perp}T_{e}+\left(  \chi_{P}-\mu_{P}\right)  T_{e}%
\nabla_{\perp}n_{e}\right]  }{m_{e}\Omega_{e}^{2}},\nonumber\\
\vec{q}_{eH}  &  \equiv\frac{5T_{e}\hat{b}\times\left[  \chi_{H}n_{e}%
\nabla_{\perp}T_{e}+\left(  \chi_{H}-\lambda\right)  T_{e}\nabla_{\perp}%
n_{e}\right]  }{m_{e}\Omega_{e}},\nonumber\\
\vec{q}_{e\parallel}  &  \equiv\frac{5T_{e}\left[  \chi_{\parallel}n_{e}%
\nabla_{\parallel}T_{e}+\left(  \chi_{\parallel}-\mu_{\parallel}\right)
T_{e}\nabla_{\parallel}n_{e}\right]  }{m_{e}\left\langle \nu_{e}\right\rangle
_{e}}, \label{q_e_explicit}%
\end{align}%
\begin{equation}
\chi_{P}\equiv2\mu_{P}+\alpha_{P}-\beta_{P}-\rho_{P},\qquad\chi_{H}%
\equiv2\lambda-1,\qquad\chi_{\parallel}\equiv2\mu_{\parallel}-\beta
_{\parallel}, \label{xis}%
\end{equation}

In addition to $\lambda$ defined by (\ref{lambda}),
Eqs.~(\ref{final_momentum_V})--(\ref{xis}) include
\begin{equation}
\alpha_{P}\equiv\frac{\int_{0}^{\infty}\frac{d\left(  v^{3}\nu_{e}%
(v_{e})\right)  }{dv_{e}}\ F_{0}dv_{e}}{3\int_{0}^{\infty}\nu_{e}(v_{e}%
)F_{0}v_{e}^{2}dv_{e}},\qquad\alpha_{\parallel}\equiv\frac{\int_{0}^{\infty
}\frac{d\left(  v_{e}^{3}/\nu_{e}(v_{e})\right)  }{dv}\ F_{0}dv_{e}}{3\int%
_{0}^{\infty}\frac{1}{\nu_{e}(v_{e})}F_{0}v^{2}dv_{e}}, \label{alphy_again}%
\end{equation}
\begin{subequations}
\label{rhos_again}%
\begin{align}
\rho_{\parallel}  &  \equiv\frac{\left(  \int_{0}^{\infty}\frac{d}{dv_{e}%
}\left(  \frac{v_{e}^{5}}{\nu_{e}(v_{e})}\right)  F_{0}dv_{e}\right)  \left(
\int_{0}^{\infty}v_{e}^{2}F_{0}dv_{e}\right)  }{5\left(  \int_{0}^{\infty
}v_{e}^{4}F_{0}dv_{e}\right)  \left(  \int_{0}^{\infty}\frac{v_{e}^{2}}%
{\nu_{e}(v_{e})}F_{0}dv_{e}\right)  },\label{rho_II_again}\\
\rho_{P}  &  \equiv\frac{\left(  \int_{0}^{\infty}\frac{d}{dv_{e}}\left(
\nu_{e}(v_{e})v_{e}^{5}\right)  F_{0}dv\right)  \left(  \int_{0}^{\infty}%
v_{e}^{2}F_{0}dv_{e}\right)  }{5\left(  \int_{0}^{\infty}v_{e}^{4}F_{0}%
dv_{e}\right)  \left(  \int_{0}^{\infty}\nu_{e}(v_{e})v_{e}^{2}F_{0}%
dv_{e}\right)  }, \label{rho_perp_again}%
\end{align}
\end{subequations}
\begin{subequations}
\label{betas}%
\begin{align}
\beta_{P}  &  \equiv\frac{\left(  \int_{0}^{\infty}\nu_{e}(v_{e})F_{0}%
v_{e}^{4}dv_{e}\right)  \int_{0}^{\infty}F_{0}v_{e}^{2}dv_{e}}{\left(
\int_{0}^{\infty}\nu_{e}(v_{e})F_{0}v_{e}^{2}dv_{e}\right)  \int_{0}^{\infty
}F_{0}v_{e}^{4}dv_{e}},\label{beta_P}\\
\beta_{\parallel}  &  \equiv\frac{\left(  \int_{0}^{\infty}\frac{F_{0}%
v_{e}^{4}}{\nu_{e}(v_{e})}\ dv_{e}\right)  \int_{0}^{\infty}F_{0}v_{e}%
^{2}dv_{e}}{\left(  \int_{0}^{\infty}\frac{F_{0}v^{2}}{\nu_{e}(v_{e})}%
\ dv_{e}\right)  \int_{0}^{\infty}F_{0}v_{e}^{4}dv_{e}}, \label{beta_II}%
\end{align}%
\end{subequations}
\begin{equation}
\delta_{P}\equiv\frac{\left(  \int_{0}^{\infty}\nu_{e}^{2}F_{0}v_{e}^{4}%
dv_{e}\right)  \left(  \int_{0}^{\infty}F_{0}v_{e}^{2}dv_{e}\right)  ^{2}%
}{\left(  \int_{0}^{\infty}\nu_{e}F_{0}v_{e}^{2}dv_{e}\right)  ^{2}\left(
\int_{0}^{\infty}F_{0}v_{e}^{4}dv_{e}\right)  }, \label{delta_P}%
\end{equation}
\begin{subequations}
\label{mus}%
\begin{align}
\mu_{P}  &  \equiv\frac{3\left(  \int_{0}^{\infty}\nu_{e}(v_{e})F_{0}v_{e}%
^{6}dv_{e}\right)  \left(  \int_{0}^{\infty}F_{0}v_{e}^{2}dv_{e}\right)  ^{2}%
}{5\left(  \int_{0}^{\infty}\nu_{e}(v_{e})F_{0}v_{e}^{2}dv_{e}\right)  \left(
\int_{0}^{\infty}F_{0}v_{e}^{4}dv_{e}\right)  ^{2}},\label{mu_P}\\
\mu_{\parallel}  &  \equiv\frac{3\left(  \int_{0}^{\infty}\frac{F_{0}v^{6}%
}{\nu_{e}(v_{e})}\ dv_{e}\right)  \left(  \int_{0}^{\infty}F_{0}v_{e}%
^{2}dv_{e}\right)  ^{2}}{5\left(  \int_{0}^{\infty}\frac{F_{0}v^{2}}{\nu
_{e}(v_{e})}\ dv_{e}\right)  \left(  \int_{0}^{\infty}F_{0}v_{e}^{4}%
dv_{e}\right)  ^{2}}, \label{mu_II}%
\end{align}%
\end{subequations}
\begin{equation}
\xi\equiv\left\langle \frac{1}{\nu_{e}}\right\rangle _{e}\left\langle \nu
_{e}\right\rangle _{e}=\frac{\left(  \int_{0}^{\infty}\nu_{e}(v_{e}%
)F_{0}(v_{e})v_{e}^{2}dv_{e}\right)  \left(  \int_{0}^{\infty}\nu_{e}%
^{-1}(v_{e})F_{0}(v_{e})v_{e}^{2}dv_{e}\right)  }{\left(  \int_{0}^{\infty
}F_{0}(v_{e})v_{e}^{2}dv_{e}\right)  ^{2}}. \label{xi}%
\end{equation}

For constant $\nu_{e}$ and arbitrary $F_{0}(v)$, we have $\alpha_{P,\parallel
}=\rho_{P,\parallel}=\beta_{P,\parallel}=\delta_{P}=\xi=1$, $\mu
_{P,\parallel}=\lambda$, $\chi_{P,H,\parallel}=2\lambda-1$ and $\mathbf{M}%
=\mathbf{N}$, so that Eqs.~(\ref{Thermal_balance_fin})--(\ref{q_e_explicit})
reduce to Eqs.~(\ref{el_temper_constant_final})--(\ref{q_IIPH_stac}). If,
additionally, $F_{0}(v)$ is Maxwellian then we have even simpler parameters:
$\mu_{P,\parallel}=\chi_{P,H,\parallel}=\lambda=1$.

\begin{figure}
\begin{center}
\includegraphics[width=17cm]{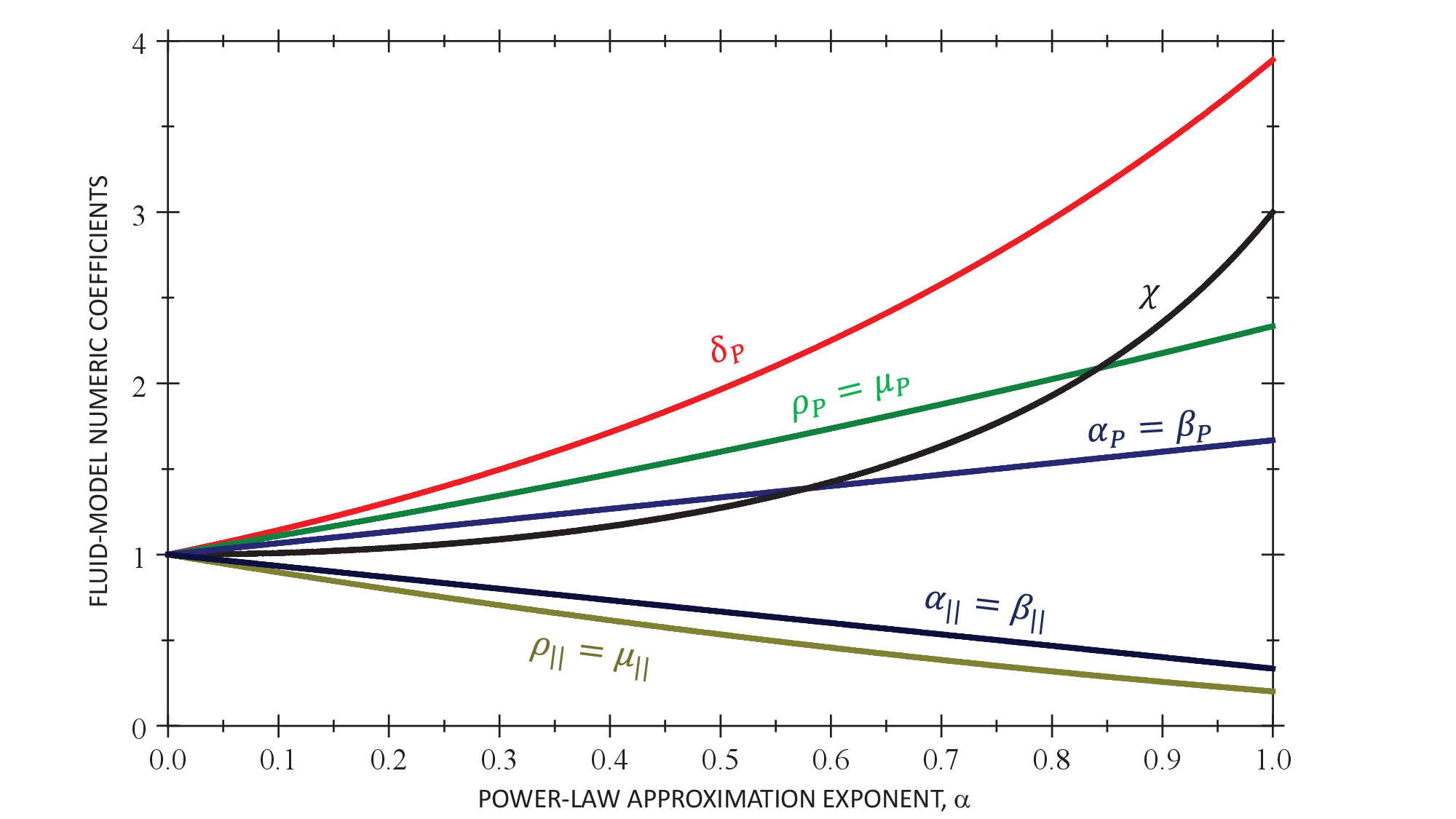}
\end{center}
\caption{Fluid-model coefficients for the power-law dependent \emph{e-n} collision frequency,
$\nu_{en}(v_e)\propto v_e^{2\alpha}$.}
\label{fig:1}
\end{figure}

In a broad range of electron energies, $\mathcal{E}_{e}\lesssim0.3$~eV,
the velocity dependence of $\nu_{e}$ in the lower ionosphere can be
approximated by a simple power-law dependence, $\nu_{e}\propto v_{e}^{2\alpha
}$, with $\alpha\approx5/6$ [\citep{Gurevich:Nonlinear78}, Sect. 2.3.1, Fig. 7]
or, practically to the same accuracy, with $\alpha=1$ \citep{Dimant:Kinetic95a}%
. For the general power-law dependent $\nu_{e}\propto v_{e}^{2\alpha}$ with
$\alpha$ in the range between $0$ and $1$ and Maxwellian $F_{0}(v_{e})$,
Eqs.~(\ref{alphy_again})--(\ref{xi}) simplify dramatically,%
\[
\alpha_{P}=\beta_{P}=1+\frac{2\alpha}{3},\qquad\alpha_{\parallel}%
=\beta_{\parallel}=1-\frac{2\alpha}{3},
\]%
\begin{equation}
\rho_{\parallel}=\mu_{\parallel}=\frac{\left(  3-2\alpha\right)  \left(
5-2\alpha\right)  }{15},\qquad\rho_{P}=\mu_{P}=\frac{\left(  3+2\alpha\right)
\left(  5+2\alpha\right)  }{15}, \label{simplify_dram}%
\end{equation}%
\[
\delta_{P}=\frac{\sqrt{\pi}\Gamma\left(  5/2+2\alpha\right)  }{3\Gamma
^{2}\left(  3/2+\alpha\right)  },\qquad\xi=\frac{1-4\alpha^{2}}{\sin\frac
{\pi\left(  2\alpha+1\right)  }{2}}=\frac{4\alpha^{2}-1}{\sin\frac{\pi\left(
2\alpha-1\right)  }{2}}.
\]
The case of $\alpha=1/2$ corresponds to hard-sphere collisions. In this case,
the indeterminate expression for $\xi$ yields $4/\pi\approx1.273$. For
$\alpha=5/6$ \citep{Gurevich:Nonlinear78}, we have $\alpha_{P}=\beta_{P}%
\approx1.556$, $\alpha_{\parallel}=\beta_{\parallel}\approx0.444$,
$\rho_{\parallel}=\mu_{\parallel}\approx0.296$, $\rho_{P}=\mu_{P}\approx
2.074$, $\delta_{P}\approx3.095,$ and $\xi\approx2.053$. For $\alpha=1$
\citep{Dimant:Kinetic95a}, all these factors deviate from unity even further,
e.g., $\alpha_{\parallel}=\beta_{\parallel}\approx0.333$, $\delta
_{P}\approx3.889$, and $\xi=3$. Thus the quantitative effect of the velocity
dependence of $\nu_{e}(v)$ is significant and should not be ignored.
Figure~\ref{fig:1} shows the coefficients given by Eq.~(\ref{simplify_dram}) for general
values of the power-law exponent $\alpha$ within the physically realistic
range of $0\leq\alpha\leq1$.

Comparison of general energy-balance Eq.~(\ref{Thermal_balance_fin}) with
Eq.~(\ref{el_temper_constant_final}) shows that the velocity dependence of the
collision parameters results not only in more complicated heat conductivity,
frictional heating, and cooling, but in additional terms associated with the
plasma motion and gradients in the parallel to $\vec{B}$ direction. It is
important that these seemingly collisionless terms originate fully from
electron-neutral collisions due to the velocity distribution of $\nu_{e}(v)$.
Similar effects in the Hall and Pedersen directions are inconsequential and
are neglected here. One should not, however, neglect Hall and Pedersen
components in the heat conductivity because the plasma temperature and density
gradients in those directions can be much sharper than those in the parallel direction.

Now we discuss the last (cooling) term in the RHS\ of
Eq.~(\ref{Thermal_balance_fin}). For general velocity-dependent $\delta
_{en}(v),$ but a Maxwellian distribution function $F_{0}\propto\exp
[-m_{e}v^{2}/(2T_{e})]$, using (\ref{nTV}), we reduce this term to%
\begin{equation}
-2\pi m_{e}\int_{0}^{\infty}\!\!v_{e}^{3}\delta_{en}\nu_{e}\left(  v_{e}%
F_{0}+\frac{T_{n}}{m_{e}}\frac{dF_{0}}{dv_{e}}\right)  dv_{e}=\frac{3}%
{2}\ \frac{\left\langle \delta_{en}v_{e}^{2}\nu_{e}\right\rangle _{e}n_{e}%
}{\left\langle v_{e}^{2}\right\rangle _{e}}\left(  T_{n}-T_{e}\right)  .
\label{for_Max}%
\end{equation}
For general $F_{0}$, but constant $\delta_{en}$, we can rewrite the cooling
term in Eq.~(\ref{Thermal_balance_fin}) as
\begin{equation}
-2\pi m_{e}\int_{0}^{\infty}\!\!v_{e}^{3}\delta_{en}\nu_{e}\left(  v_{e}%
F_{0}+\frac{T_{n}}{m_{e}}\frac{dF_{0}}{dv_{e}}\right)  dv_{e}=\frac{3}%
{2}\ \delta_{en}\left\langle \nu_{e}\right\rangle _{e}n_{e}\left(  \alpha
_{P}T_{n}-\beta_{P}T_{e}\right)  , \label{cooling}%
\end{equation}
where we integrated by parts and used~Eqs.~(\ref{alphy_again}) and
(\ref{beta_P}). Equation~(\ref{cooling}) shows that for general non-Maxwellian
$F_{0}$ the cooling term is not necessarily proportional to the temperature
difference $(T_{n}-T_{e})$. However, for the power-law dependent $\nu
_{e}\propto v_{e}^{2\alpha}$ and Maxwellian $F_{0}$, according to
Eq.~(\ref{simplify_dram}), we have $\beta_{P}=\alpha_{P}=1+2\alpha/3$. In this
case, the structure of the cooling term proportional to $\alpha_{P}%
\left\langle \nu_{e}\right\rangle n_{e}$ matches that of the frictional
heating term for a purely perpendicular field, $\alpha_{P}\left\langle \nu
_{e}\right\rangle m_{e}n_{e}V_{e\perp}^{2}$, as seen from the first term in
the RHS of Eq.~(\ref{Thermal_balance_fin}).

\section{Discussion}
\label{Discussion}

When applying a fluid model, either for analytic calculations or simulations,
it is important to have the corresponding equations with accurate parameters
applicable to the relevant physical conditions. These equations and parameters
are usually derived from the kinetic theory, so that their accuracy is determined by the
accuracy of the underlying kinetic approach.

Based on two different approximate kinetic approaches, this paper derives the fluid-model equations that
describe low-frequency plasma processes in the highly dissipative E-region
ionosphere. The treatment is restricted to collisions of the plasma
particles, ions or electrons, with the neutral molecules only; no Coulomb
collisions are considered. The neglect of Coulomb collisions at the E-region
ionosphere is usually well justified, although sometimes electron-electron collisions
may play a role, resulting in a more efficient maxwellization of the electron distribution
function \citep{Dimant:Kinetic95a}. Such maxwellization makes the fluid model
(as opposed to the pure kinetic theory) even more applicable. For the plasma particle collisions with
neutrals (elastic or inelastic), we assume the known cross-sections as functions of
the colliding particle velocities. These velocity dependencies of the collisional cross-sections can be taken from the literature
\citep[e.g., for the dominant electron-nitrogen collisions, see][and references therein]{Itikawa:Cross06,Song:Cross23}.
Assuming these known cross-sections, we can always calculate the velocity dependencies of the kinetic
collision frequencies, $\nu_p(\vec{v}_p)$ ($p=i,e,n$) that are used in this paper.

The resultant fluid-model parameters are expressed in general integral forms through the known
velocity dependencies of $\nu_p$. For the most important plasma processes, such as the small- to medium-scale
cross-field plasma instabilities (the thermal Farley-Buneman and
gradient drift instabilities), closed 5-moment multi-fluid
models are usually sufficient for the accurate fluid description. Given the
plasma species $p$, the 5-moment set of the unknowns includes
the particle density ($n_{p}$), temperature ($T_{p}$), and the three
components of the mean drift velocity, $\vec{V}_{p}$.

For the ionospheric ions ($p=i$), we have employed the well-known (and fairly simple)
BGK collisional model. For the heavy ions, the applicability of the BGK
collision operator can be justified by the fact that within the thermal bulk
and around, the kinetic ion-neutral collision frequency $\nu_{in}$ is
approximately constant, i.e., velocity-independent; this approximation corresponds to so-called Maxwell molecule
collisions \citep{Schunk:Ionospheres09}. Additionally, in the E-region the ion masses ionosphere are fairly
close to the neutral-molecule masses, $m_{i}\simeq m_{n}$. As we demonstrate
in section~\ref{Section: for ions}, in the case of $m_{i}=m_{n}$ the
oversimplified BGK model results even in a quantitatively accurate frictional
heating and cooling terms, see the RHS of equation~(\ref{ion_temperature_equation}).
We should bear in mind, however,
that under sufficiently strong electric field, $E\gtrsim m_{i}\nu_{i}V_{Ti}/e$,
the ion distribution function can be significantly distorted with an
appreciable deviation from Maxwellian. Although the major ion-fluid terms in these cases
remain valid, the entire 5-moment model cannot necessarily be easily closed,
and hence its validity may be questionable. The factor of strong electric field
is usually of importance for the high-latitude E region under conditions of
severe magnetospheric perturbations (geomagnetic storms or substorms), while
at the equatorial E region the electric fields are typically much weaker, so
that the closed 5-moment ion fluid model is usually more applicable.

The central part of this paper is the derivation of the 5-moment fluid
equations for electrons ($p=e$). For the light electrons, unlike the ions, the simple
BGK model cannot serve even as a crude approximation. As we explained in
section~\ref{Section: for electrons}, the reasons for the total BGK-model
inapplicability are the two major facts: (1) the mean rate of the collisional
loss of the electron energy is much less than the corresponding loss of the
electron momentum, so that the electron behavior cannot be described by a
single collisional parameter; (2) the kinetic collisional frequency
has a pronounced electron velocity dependence,  $\nu_{e}(v_e)$. The first fact leads to a strong
isotropization of the electron velocity distribution in the velocity space, while the speed
dependence of the electron velocity distribution is effectively decoupled from
its angular dependence. The second fact leads to
noticeable modifications of the electron fluid coefficients and even to
occurrence of additional thermal-diffusion terms. As a result, in the general
case the fluid-model coefficients acquire additional dimensionless multipliers whose values are determined by some
integral relations over the entire electron distribution function, see
Eqs.~(\ref{qq_e})--(\ref{xi}). For the Maxwellian function, and especially for the power-law
dependencies of the $\nu_{e}$-speed dependence, $\nu_{e}(v_{e})\propto
v_{e}^{2\alpha}$, these general integral relations reduce to simple algebraic
ones, see Eq.~(\ref{simplify_dram}) and Figure~\ref{fig:1}. From that
figure we see that some numerical multipliers can deviate from the unity quite significantly,
although they still remain in the same order of magnitude. A better quantitative knowledge of these
fluid coefficients is important for accurate calculations and predictions of
physical characteristics of various plasma wave and other processes. As our future knowledge of
the speed dependence of the kinetic collision rates becomes more precise,
using the more general integral relationships obtained here one can
improved the values of the corresponding fluid-model coefficients.

The kinetic approach for the electron description employed in this paper is based on the expansion of the
electron velocity distribution in Legendre polynomials in the velocity space and keeping the two first
terms of such expansion, see Eq.(\ref{f_approx}): the dominant isotropic part, $F_{0}(\vec{r},t,v_{e})$,
and a small directional part, $|\vec{f}_{1}(\vec{r},t,v_{e})|\cos\theta$,
where $\theta$ is the angle between $\vec{f}_{1}$.
This approach is analogous to that employed by \citet{Gurevich:Nonlinear78}, see also \citet{Dimant:Kinetic95a}, although, as we
explained in section~\ref{General kinetic approach for electrons}, Gurevich's fluid
equations for electrons, \citet[][Chapter 5]{Gurevich:Nonlinear78}, derived through this kinetic approach,
differ from ours. Gurevich's equations are written in a form that does not include explicit adiabatic and
frictional heating terms. Also, as wee mentioned above,
the ``adiabatic'' terms proportional to $(\gamma_{e,i}-1)$ in Gurevich's Eqs.~(5.3) and (5.4) are extraneous,
while their correct equivalent has already been implicitly distributed within the other terms
in Gurevich's Eqs.~(5.8)--(5.11) (and hence included twice).

An alternative kinetic approach to build an electron fluid description is based on Grad's method
\citep{Kissack:Electron95,Kissack:Effect97, Kissack:Thermal_I_08,Kissack:Thermal_II_08}.
The latter assumes a finite number of parameters that characterize the velocity distribution, implying
that the electron velocity distribution is reasonably close to
Maxwellian. Our approach is much more general in terms of the $v_{e}$-dependence,
but it restricts the angular distribution of the electron
distribution function to the simplest linear deviation from the isotropy (see above). This
approximation allows calculating vector fluxes like $n_{e}\vec{V}_{e}$ or
energy fluxes (see below), but higher-order tensor characteristics like an
anisotropic pressure, etc., may require an accuracy beyond its field of
applicability. Note, however, that high-order tensor characteristics for
electrons are not expected to be significant due to the relatively high rate
of electron distribution function isotropization associated with a small value
of $\delta_{en}\sim(2$--$4)\times10^{-3}$ within the low-energy electron
energy range, $\mathcal{E}_{e}\equiv m_{e}v_{e}^{2}/2<2\mathrm{\ }$eV
\citep{Gurevich:Nonlinear78}. Note also that under physical conditions when the
two methods are applicable both techniques actually provide reasonable close
quantitative results. At the same time, our kinetic approach provides much
simpler, and hence much more practical, algebraic expressions applicable to various
small- and medium-scale E-region processes.

\section{Conclusion}
\label{Conclusion}
Based on relevant physical conditions, we have derived improved fluid equations for the
E-region ionosphere. In this derivation, for the E-region ions and electrons we have used 
two different approximate kinetic approaches.

For the ions, we have employed the simple BGK collision operator (section~\ref{Section: for ions}).
This resulted in full 5-moment set of the continuity, momentum, and energy balance equations, 
see Eqs.~(\ref{ion_continuity}), (\ref{momentum_i}),
and (\ref{ion_temperature_equation}). Though these equations look conventional,
our derivation has demonstrated that for the E-region ions with almost equal masses of the ions
and neutrals, the BGK collision operator leads to the quantitatively accurate frictional and cooling rates.

The central part of this paper is the derivation of the electron fluid equations. For the electrons,
the BGK collisional operator is inapplicable and we have employed the kinetic approach based
on the expansion of the electron distribution function, $f_e(\vec{v}_e)$, in Legendre polynomials over
the angles in the velocity space. Due to physical conditions resulting in efficient isotropization
of $f_e(\vec{v}_e)$ allowed us to restrict the treatment to the two highest terms of the
Legendre expansion: the dominant isotropic part, $F_0(v_e)$ and a small directional part
$\vec{f_1}\cdot \vec{v}_e/v_e$. The former is responsible for calculating the scalar
fluid quantities, such as the electron density and temperature, while the directional part
allows one to calculate the electron flux (or electric current) density. An important factor in our derivations is the fact
that the electron-neutral collisional frequencies are strongly velocity-dependent. Assuming these velocity
dependencies to be known, we have derived the full set of the 5-moment equations: the continuity
equation, the momentum equation, and the thermal balance equation. 
Since the E-region electrons in all relevant low-frequency processes are
essentially inertialess, the momentum equation reduces to an explicit expression for the
electron mean drift velocity given by Eq.~(\ref{final_momentum_V}). The most non-trivial result is the thermal 
balance equation given in the general case by Eq.~(\ref{Thermal_balance_fin}), where the 
parameters are given by Eqs.~(\ref{lambda}), (\ref{alphy_again})--(\ref{xis}). For the Maxwellian 
distribution function and the power-law speed dependence of the electron-neutral collision frequency, $\nu_{e}(v_e)\propto v_e^{2\alpha}$,
the integral relationships for the fluid-model parameters reduce to simple algebraic expressions given by Eq.~(\ref{simplify_dram}), see also Fig.~\ref{fig:1}.

Comparison of general energy-balance Eq.~(\ref{Thermal_balance_fin}) with the 
corresponding equation for the velocity-independent electron collision frequency, see
Eq.~(\ref{el_temper_constant_final}), shows that the velocity dependence of the
collision parameters results in a more complicated heat conductivity,
frictional heating, and cooling, as well as in additional terms associated with the
plasma motion and gradients in the parallel to $\vec{B}$ direction. 
These terms look collisionless, but they originate exclusively from
the velocity distribution of $\nu_{e}(v)$. Similar effects in the Hall and Pedersen directions are inconsequential and
have been neglected. However, one should not neglect the Hall and Pedersen
components in the heat conductivity because the plasma temperature and density
gradients in those directions can be much sharper than those in the parallel direction.

The results of this paper could be used for a routine practical analysis when
working with actual data. The improved equations can also serve as a basis for
more accurate plasma fluid computer simulations. In the general case, the applicability of the closed 5-moment equations 
is restricted my reasonably moderate conditions of the equatorial E region. For the high-latitude E-region ionosphere, the accurate 
description may require using a fully kinetic treatment.

\section*{Acknowledgment}
The author is deeply grateful to Prof. Oppenheim for his long-term support, extensive help,
and multiple discussions. This work was funded by NSF Grants 1755350, 1007789, 0442075
and NASA Grants 80NSSC21K1322, 80NSSC19K0080.

\bigskip

\appendix\noindent{\textbf{{\Large APPENDIX}}}

\setcounter{equation}{0} \renewcommand{\theequation}{A.\arabic{equation}}

\section*{Moment equations for velocity-dependent $\nu_{e}(v_{e})$}

In this Appendix, we derive Eqs.~(\ref{final_momentum_V}) and
(\ref{Thermal_balance_fin}). The key step of this derivation is expressing the
electric field, $\vec{E}$, in terms of the electron average drift velocity,
$\vec{V}_{e}$, and the $n_{e}$,$T_{e}$-gradients.

For velocity-dependent $\nu_{e}(v)$, which cannot be factored out of the
corresponding integrals, we use the following approach. Using
Eq.~(\ref{f_1_nach}) and integrating by parts, we obtain from Eq.~(\ref{nTV})
\begin{equation}
\vec{V}_{e}=-\ \frac{4\pi}{3n_{e}}\int_{0}^{\infty}v^{4}\mathbf{N}\cdot\nabla
F_{0}dv-\mathbf{M}\cdot\frac{e\vec{E}}{m_{e}}, \label{V_e_povtor}%
\end{equation}
where $\mathbf{M}$ is given by Eq.~(\ref{M_e_again}) with $\alpha
_{P,\parallel}$ defined by Eq.~(\ref{alphy_again}). By inverting the matrix
$\mathbf{M}$, we obtain%
\begin{equation}
\frac{e\vec{E}}{m_{e}}=-\mathbf{M}^{-1}\cdot\left(  \vec{V}_{e}+\frac{4\pi
}{3n_{e}}\int_{0}^{\infty}v^{4}\mathbf{N}(v)\cdot\nabla F_{0}dv\right)  .
\label{E_again}%
\end{equation}
Using Eq.~(\ref{f_1_nach}), we obtain%
\begin{equation}
\vec{f}_{1}(u)=-\mathbf{N}(u)\cdot\left[  u\nabla F_{0}+\mathbf{M}^{-1}%
\cdot\left(  \vec{V}_{e}+\frac{4\pi}{3n_{e}}\int_{0}^{\infty}\mathbf{N}%
(v)\cdot\nabla F_{0}(v)v^{4}dv\right)  \frac{\partial F_{0}}{\partial
u}\right]  . \label{f_1_polnoe}%
\end{equation}
To complete the derivation of the momentum and temperature-balance equations
using Eqs.~(\ref{c_nachala}) and (\ref{equivoka}), we need to calculate there
the two integrals involving $\vec{f}_{1}$. Both integrals have the same
structure described by $\int_{0}^{\infty}\vec{f}_{1}g(v)v^{3}dv$, where the
function $g(v)$ is either $\nu_{e}(v)$ or $v^{2}$. Using integration by parts,
changing the variables, and re-ordering the integration wherever needed, we
obtain%
\begin{align*}
&  \int_{0}^{\infty}\vec{f}_{1}g(v_{e})v_{e}^{3}dv_{e}=\int_{0}^{\infty}%
\frac{d\left(  g(u)u^{3}\mathbf{N}(u)\cdot\mathbf{M}^{-1}\cdot\vec{V}%
_{e}\right)  }{du}\ F_{0}(u)du\\
&  +\int_{0}^{\infty}\left[  \frac{4\pi}{3n_{e}}\int_{0}^{\infty}%
\frac{d\left(  g(u)u^{3}\mathbf{N}(u)\cdot\mathbf{M}^{-1}\right)  }{du}%
\ F_{0}(u)du-g(v)\mathbf{I}\right] \\
&  \cdot\mathbf{N}(v)\cdot\nabla F_{0}(v_{e})v_{e}^{4}dv_{e},
\end{align*}
where $\mathbf{I}$ is the unit tensor: $\mathbf{I}_{\alpha\beta}=1$ if
$\alpha=\beta$ and $\mathbf{I}_{\alpha\beta}=0$ otherwise. As a result, for
the integrals in Eqs.~(\ref{c_nachala}) and (\ref{equivoka}) we obtain%
\begin{align}
&  \frac{4\pi m_{e}}{3}\int_{0}^{\infty}\nu_{e}\vec{f}_{1}v_{e}^{3}%
dv_{e}\nonumber\\
&  =\frac{4\pi m_{e}}{3}\int_{0}^{\infty}\frac{d\left(  \nu_{e}(u)u^{3}%
\mathbf{N}(u)\cdot\mathbf{M}^{-1}\cdot\vec{V}_{e}\right)  }{du}F_{0}%
(u)du\nonumber\\
&  +\frac{4\pi m_{e}}{3}\int_{0}^{\infty}\left[  \frac{4\pi}{3n_{e}}\int%
_{0}^{\infty}\frac{d\left(  \nu_{e}(u)u^{3}\mathbf{N}(u)\cdot\mathbf{M}%
^{-1}\right)  }{du}F_{0}(u)du-\nu_{e}(v)\mathbf{I}\right]
\label{int_f_1_nu_v3_dv}\\
&  \cdot\mathbf{N}(v)\cdot\nabla F_{0}(v_{e})v_{e}^{4}dv_{e},\nonumber
\end{align}
and%
\begin{align}
&  \frac{2\pi m_{e}}{3}\ \nabla\cdot\int_{0}^{\infty}\ \vec{f}_{1}v_{e}%
^{5}dv_{e}=\frac{2\pi m_{e}}{3}\ \nabla\cdot\int_{0}^{\infty}\frac{d\left(
u^{5}\mathbf{N}(u)\cdot\mathbf{M}^{-1}\cdot\vec{V}_{e}\right)  }{du}%
F_{0}(u)du\nonumber\\
&  +\frac{2\pi m_{e}}{3}\ \nabla\cdot\int_{0}^{\infty}\left[  \frac{4\pi
}{3n_{e}}\int_{0}^{\infty}\frac{d\left(  u^{5}\mathbf{N}(u)\cdot
\mathbf{M}^{-1}\right)  }{du}F_{0}(u)du-v^{2}\mathbf{I}\right]
\label{int_f_1v5_dv}\\
&  \cdot\mathbf{N}(v)\cdot\nabla F_{0}(v_{e})v_{e}^{4}dv_{e}.\nonumber
\end{align}

First, we calculate the integrals in the RHS of Eq.~(\ref{int_f_1_nu_v3_dv}).
After a simple algebra, we obtain%
\begin{align}
&  \frac{4\pi}{3n_{e}}\int_{0}^{\infty}\frac{d\left(  \nu_{e}(u)u^{3}%
\mathbf{N}(u)\cdot\mathbf{M}^{-1}\cdot\vec{V}_{e}\right)  }{du}\ F_{0}%
(u)du\nonumber\\
&  \approx\alpha_{P}\left\langle \nu_{e}\right\rangle \vec{V}_{e\perp}%
+\frac{\left(  \alpha_{P}^{2}-\mu\right)  \left\langle \nu_{e}\right\rangle
^{2}}{\Omega_{e}}\left(  \hat{b}\times\vec{V}_{e\perp}\right)  +\frac
{\left\langle \nu_{e}\right\rangle \vec{V}_{\parallel}}{\alpha_{\parallel}\xi},
\label{new_frict}%
\end{align}
where
\begin{equation}
\mu\equiv\frac{\left(  \int_{0}^{\infty}\frac{d\left(  \nu_{e}^{2}(v_{e}%
)v_{e}^{3}\right)  }{dv_{e}}\ F_{0}dv_{e}\right)  \int_{0}^{\infty}F_{0}%
v_{e}^{2}dv_{e}}{3\left(  \int_{0}^{\infty}\nu_{e}v_{e}^{2}F_{0}dv_{e}\right)
^{2}}, \label{mu}%
\end{equation}%
and similar for the second term in the RHS of (\ref{int_f_1v5_dv}),%
\begin{align}
&  \frac{4\pi}{3n_{e}}\int_{0}^{\infty}\nu_{e}\vec{f}_{1}v_{e}^{3}%
dv_{e}\approx\alpha_{P}\left\langle \nu_{e}\right\rangle \vec{V}_{e\perp
}+\frac{\left(  \alpha_{P}^{2}-\mu\right)  \left\langle \nu_{e}\right\rangle
^{2}}{\Omega_{e}}\left(  \hat{b}\times\vec{V}_{e\perp}\right)  +\frac
{\left\langle \nu_{e}\right\rangle \vec{V}_{\parallel}}{\alpha_{\parallel}\xi
}\nonumber\\
&  -\ \frac{\left[  \alpha_{P}\left(  \alpha_{P}-\beta_{P}\right)  -\mu
+\delta_{P}\right]  \left\langle \nu_{e}\right\rangle ^{2}\nabla_{\perp
}\left(  n_{e}T_{e}\right)  }{\Omega_{e}^{2}m_{e}n_{e}}\label{ves_pervyj}\\
&  +\frac{\left(  \alpha_{P}-\beta_{P}\right)  \left\langle \nu_{e}%
\right\rangle \hat{b}\times\nabla_{\perp}\left(  n_{e}T_{e}\right)  }%
{\Omega_{e}m_{e}n_{e}}+\left(  \frac{\beta_{\parallel}}{\alpha_{\parallel}%
}-1\right)  \frac{\nabla_{\parallel}\left(  n_{e}T_{e}\right)  }{m_{e}n_{e}%
}.\nonumber
\end{align}
Here and below, all parameters have been defined by Eqs.~(\ref{alphy_again}%
)--(\ref{xi}). Substituting Eq.~(\ref{ves_pervyj}) to Eq.~(\ref{c_nachala})
and neglecting the second-order small terms $\sim\nu_{e}^{2}/\Omega_{e}^{2}$
wherever applicable, we obtain%
\begin{equation}
\frac{e\vec{E}}{m_{e}}-\Omega_{e}\hat{b}\times\vec{V}_{e\perp}+\left\langle
\nu_{e}\right\rangle _{e}\left(  \alpha_{P}\vec{V}_{e\perp}+\frac{\vec
{V}_{e\parallel}}{\alpha_{\parallel}\xi}\right)  +\frac{\nabla_{\perp}\left(
n_{e}T_{e}\right)  }{m_{e}n_{e}}+\frac{\beta_{\parallel}}{\alpha_{\parallel}%
}\frac{\nabla_{\parallel}\left(  n_{e}T_{e}\right)  }{m_{e}n_{e}}=0.
\label{full_the _same}%
\end{equation}
Expressing the combination of the second and third terms in the LHS\ of
Eq.~(\ref{full_the _same}) as a matrix-by-vector product,%
\[
-\Omega_{e}\hat{b}\times\vec{V}_{e\perp}+\left\langle \nu_{e}\right\rangle
_{e}\left(  \alpha_{P}\vec{V}_{e\perp}+\frac{\vec{V}_{e\parallel}}%
{\alpha_{\parallel}\xi}\right)  =\left[
\begin{array}
[c]{ccc}%
\alpha_{P}\left\langle \nu_{e}\right\rangle _{e} & \Omega_{e} & 0\\
-\Omega_{e} & \alpha_{P}\left\langle \nu_{e}\right\rangle _{e} & 0\\
0 & 0 & \frac{\left\langle \nu_{e}\right\rangle _{e}}{\alpha_{\parallel}\xi}%
\end{array}
\right]  \vec{V}_{e},
\]
and resolving the resultant matrix equation for $\vec{V}_{e}$ to the
first-order accuracy with respect to $\nu_{e}/\Omega_{e}$,%
\[
\left[
\begin{array}
[c]{ccc}%
\alpha_{P}\left\langle \nu_{e}\right\rangle _{e} & \Omega_{e} & 0\\
-\Omega_{e} & \alpha_{P}\left\langle \nu_{e}\right\rangle _{e} & 0\\
0 & 0 & \frac{\left\langle \nu_{e}\right\rangle _{e}}{\alpha_{\parallel}\xi}%
\end{array}
\right]  ^{-1}\approx\left[
\begin{array}
[c]{ccc}%
\frac{\alpha_{P}\left\langle \nu_{e}\right\rangle _{e}}{\Omega_{e}^{2}} &
-\ \frac{1}{\Omega_{e}} & 0\\
\frac{1}{\Omega_{e}} & \frac{\alpha_{P}\left\langle \nu_{e}\right\rangle _{e}%
}{\Omega_{e}^{2}} & 0\\
0 & 0 & \frac{\alpha_{\parallel}\xi}{\left\langle \nu_{e}\right\rangle _{e}}%
\end{array}
\right]  =\mathbf{M}\mathbf{,}%
\]
we arrive at the explicit expression for the electron fluid drift velocity
$\vec{V}_{e}$ given by Eq.~(\ref{final_momentum_V}).

Further, we calculate the integrals in Eq.~(\ref{int_f_1v5_dv}). To the
first-order accuracy with respect to $\nu_{e}/\Omega_{e}$, we obtain%
\begin{align}
&  \frac{2\pi m_{e}}{3}\ \nabla\cdot\int_{0}^{\infty}\frac{d}{dv}\left(
v_{e}^{5}\mathbf{N}\cdot\mathbf{M}^{-1}\cdot\vec{V}_{e}\right)  F_{0}%
dv_{e}\nonumber\\
&  \approx\frac{5}{2}\left\{  \nabla_{\perp}\cdot\left(  \vec{V}_{e\perp}%
n_{e}T_{e}\right)  +\nabla_{\parallel}\left(  \frac{\rho_{\parallel
}V_{e\parallel}n_{e}T_{e}}{\alpha_{\parallel}}\right)  \right. \nonumber\\
&  \left.  +\frac{\left\langle \nu_{e}\right\rangle _{e}\left(  \rho
_{P}-\alpha_{P}\right)  }{\Omega_{e}}\ \nabla_{\perp}\cdot\left[  \hat
{b}\times\left(  \vec{V}_{e\perp}n_{e}T_{e}\right)  \right]  \right\}  .
\label{prom_1}%
\end{align}
and%
\begin{align}
&  \frac{4\pi}{3n_{e}}\int_{0}^{\infty}\frac{d\left(  u^{5}\mathbf{N}%
(u)\cdot\mathbf{M}^{-1}\right)  }{du}\ F_{0}(u)du\nonumber\\
&  \approx\frac{5T_{e}}{m_{e}}\left[
\begin{array}
[c]{ccc}%
1 & \frac{\left(  \alpha_{P}-\rho_{P}\right)  \left\langle \nu_{e}%
\right\rangle _{e}}{\Omega_{e}} & 0\\
\frac{\left(  \rho_{P}-\alpha_{P}\right)  \left\langle \nu_{e}\right\rangle
_{e}}{\Omega_{e}} & 1 & 0\\
0 & 0 & \frac{\rho_{\parallel}}{\alpha_{\parallel}\left\langle 1/\nu
_{e}\right\rangle _{e}}%
\end{array}
\right]  . \label{prom_2}%
\end{align}
where the dimensionless parameters of order unity $\rho_{\parallel,P}$ are
defined by Eq.~(\ref{rhos_again}). Using Eqs.~(\ref{prom_1}) and
(\ref{prom_2}) and proceeding with calculations similar to those above, we
obtain%
\begin{align}
&  \frac{2\pi m_{e}}{3}\ \nabla\cdot\int_{0}^{\infty}\left[  \frac{4\pi
}{3n_{e}}\int_{0}^{\infty}\frac{d\left(  u^{5}\mathbf{N}(u)\cdot
\mathbf{M}^{-1}\right)  }{du}F_{0}(u)du-v^{2}\mathbf{I}\right] \nonumber\\
&  \cdot\mathbf{N}(v)\cdot\nabla F_{0}(v_{e})v_{e}^{4}dv_{e}\nonumber\\
&  =\nabla_{\perp}\cdot\frac{5\left\langle \nu_{e}\right\rangle \left[
\left(  \beta_{P}+\rho_{P}-\alpha_{P}\right)  T_{e}\nabla_{\perp}\left(
n_{e}T_{e}\right)  -\mu_{P}\nabla_{\perp}\left(  n_{e}T_{e}^{2}\right)
\right]  }{m_{e}\Omega_{e}^{2}}\nonumber\\
&  +\nabla_{\perp}\cdot\frac{5\hat{b}\times\left[  T_{e}\nabla_{\perp}\left(
n_{e}T_{e}\right)  -\lambda\nabla_{\perp}\left(  n_{e}T_{e}^{2}\right)
\right]  }{m_{e}\Omega_{e}}\nonumber\\
&  +\nabla_{\parallel}\cdot\frac{5\left[  \beta_{\parallel}T_{e}\nabla_{\perp
}\left(  n_{e}T_{e}\right)  -\mu_{\parallel}\nabla_{\perp}\left(  n_{e}%
T_{e}^{2}\right)  \right]  }{m_{e}\left\langle \nu_{e}\right\rangle }
\label{chlen_new}%
\end{align}
Separating the temperature and density gradients, after introducing linear
combinations of the numerical coefficients $\chi_{P,H,\parallel}$, as defined
in (\ref{xis}), we obtain from (\ref{prom_1}) and (\ref{chlen_new})%
\begin{align}
&  \frac{2\pi m_{e}}{3}\ \nabla\cdot\int_{0}^{\infty}\ \vec{f}_{1}v_{e}%
^{5}dv_{e}=\frac{5}{2}\left[  \nabla_{\perp}\cdot\left(  n_{e}T_{e}\vec
{V}_{e\perp}\right)  +\frac{\rho_{\parallel}}{\alpha_{\parallel}}%
\ \nabla_{\parallel}\left(  n_{e}T_{e}V_{e\parallel}\right)  \right.
\nonumber\\
&  +\left.  \left\langle \nu_{e}\right\rangle _{e}\left(  \alpha_{P}-\rho
_{P}\right)  \frac{\nabla_{\perp}\cdot\left(  \hat{b}\times n_{e}T_{e}\vec
{V}_{e\perp}\right)  }{\Omega_{e}}\right]  -\nabla\cdot\vec{q}_{e},
\label{f_1v_5dv}%
\end{align}
where parameters $\beta_{P,\parallel}$, $\mu_{P,\parallel}$ and tensors
$\mathbf{X}$, $\mathbf{\Lambda}$ are defined by (\ref{betas}), (\ref{mus}),
and (\ref{xxLL}). The last term in (\ref{f_1v_5dv}) can be written explicitly
as%
\begin{equation}
\vec{q}_{e}\equiv\vec{q}_{eP}+\vec{q}_{eH}+\vec{q}_{e\parallel}=\mathbf{X}%
\cdot\frac{\nabla T_{e}}{T_{e}}+\left(  \mathbf{X-}\mathbf{\Lambda}\right)
\cdot\frac{\nabla n_{e}}{n_{e}} \label{q_e}%
\end{equation}
where%
\begin{align}
\vec{q}_{eP}  &  \equiv\frac{5T_{e}\left\langle \nu_{e}\right\rangle \left[
\chi_{P}n_{e}\nabla_{\perp}T_{e}+\left(  \chi_{P}-\mu_{P}\right)  T_{e}%
\nabla_{\perp}n_{e}\right]  }{m_{e}\Omega_{e}^{2}}\nonumber\\
\vec{q}_{eH}  &  \equiv\frac{5T_{e}\hat{b}\times\left[  \chi_{H}n_{e}%
\nabla_{\perp}T_{e}+\left(  \chi_{H}-\lambda\right)  T_{e}\nabla_{\perp}%
n_{e}\right]  }{m_{e}\Omega_{e}}\nonumber\\
\vec{q}_{e\parallel}  &  \equiv\frac{5T_{e}\left[  \chi_{\parallel}n_{e}%
\nabla_{\parallel}T_{e}+\left(  \chi_{\parallel}-\mu_{\parallel}\right)
T_{e}\nabla_{\parallel}n_{e}\right]  }{m_{e}\left\langle \nu_{e}\right\rangle
}. \label{explicitly}%
\end{align}
From Eqs.~(\ref{equivoka}), (\ref{simple_algebra}), (\ref{ves_pervyj}),
(\ref{f_1v_5dv}), neglecting terms that are first- and second-order small with
respect to $\nu_{e}/\Omega_{e}$, we obtain thermal-balance
Eq.~(\ref{Thermal_balance_fin}).

\bibliographystyle{Frontiers-Harvard} 



\end{document}